%% file: root.tex
\documentclass[letterpaper,10 pt, conference]{ieeeconf}

\IEEEoverridecommandlockouts

\include{headers}

\begin{document}

\title{\LARGE Active Localization of Close-range Adversarial Acoustic Sources\\ for Underwater Data Center Surveillance
\vspace{-1mm}
}
\author{Adnan Abdullah, David Blow, Sara Rampazzi, and Md Jahidul Islam
%\vspace{-1mm}
\\
\small{\{adnanabdullah@, david.blow@, srampazzi@, jahid@ece.\}ufl.edu}\\
University of Florida, Gainesville, FL, USA
% <-this % stops a space
\thanks{\noindent\rule{5cm}{0.5pt}}
\thanks{This pre-print is currently under review.}
% \thanks{Corresponding author: \tt\small adnanabdullah@ufl.edu}%
}

\markboth{Journal of Oceanic Engineering,~Review.}{}

\maketitle

\input{src/00_Abstract}

\input{src/01_Intro}
\input{src/02_Related}
\input{src/03_Background}
\input{src/04_Method}

\input{src/05_Evaluation}
\input{src/06_Discussion}
\input{src/07_Conc}

\bibliographystyle{ieeetr}
\bibliography{combined}

\end{document}

%% file: headers.tex
\usepackage{cite}
\usepackage{amsmath,amssymb,amsfonts}
\usepackage{algorithmic}
\usepackage{graphicx}
\usepackage{textcomp}
\usepackage[table, svgnames, dvipsnames]{xcolor}
\usepackage{multirow}
\usepackage{soul}
\usepackage{siunitx}
\usepackage{physunits}
\usepackage{xfrac}
\usepackage{array}
\usepackage{graphicx}
\usepackage{tabularx}

\def\BibTeX{{\rm B\kern-.05em{\sc i\kern-.025em b}\kern-.08em
    T\kern-.1667em\lower.7ex\hbox{E}\kern-.125emX}}

\usepackage{xspace}
\long\def\invis#1{}

\usepackage{wrapfig}
\newcommand{\ie}{\textit{i.e.\xspace}}

\newcommand{\etal}{\textit{et al.\xspace}}

\usepackage[hyphens]{url}
\usepackage[colorlinks,bookmarksopen,bookmarksnumbered,citecolor=green,urlcolor=black]{hyperref}

\usepackage{makecell}
\long\def\invis#1{}
\usepackage{subcaption}
\usepackage[size=small]{caption}

\usepackage{hyphenat}
\usepackage{comment}
\usepackage{diagbox}
\usepackage{gensymb}

\usepackage{lipsum}
\usepackage[group-separator={,}]{siunitx}
\usepackage{tabularx}
\usepackage{balance}
\usepackage[T1]{fontenc}
\usepackage{lineno}

\usepackage{authblk}

\newcommand\rebuttal[1]{#1}

\let\labelindent\relax
\usepackage{enumitem}

%% file: src/00_Abstract.tex
\begin{abstract}
Underwater data infrastructures offer natural cooling and enhanced physical security compared to terrestrial facilities, \rebuttal{but their storage systems remain susceptible to acoustic injection attacks, where sound-induced mechanical vibrations disrupt critical I/O operations and compromise data availability.}
%\textcolor{red}{to avoid HDD vs SSD discussion, I would rather change all "HDD-based" to just "storage systems" in all the paper}\JI{good point this}
This work presents a surveillance framework for localizing and tracking such close-range adversarial acoustic sources targeting offshore infrastructures, particularly underwater data centers (UDCs). \rebuttal{We propose a scalable heterogeneous receiver configuration with one facility-mounted hydrophone and one mobile hydrophone carried by a surveillance robot. The resulting problem differs from conventional sound source localization (SSL) due to distributed facility scale, narrowband signaling with phase ambiguity, non-cooperative sources, and mobile receiver state uncertainty.}
To address these challenges, we formulate a Locus-Conditioned Maximum A-Posteriori (LC-MAP) scheme that generates acoustically informed priors, ensuring a physically plausible initial state for a joint time- and frequency-difference-of-arrival (TDOA-FDOA) filtering. We integrate this into an unscented Kalman filter (UKF) pipeline, \rebuttal{along with a multipath-aware measurement model that compensates for surface and bed reflections, and an effective measurement covariance that accounts for mobile receiver uncertainty}. Extensive Monte Carlo analyses, \rebuttal{fixed-array baseline comparisons}, Gazebo-based physics simulations, and field trials demonstrate reliable real-time localization and tracking. The framework achieves sub-meter localization accuracy and over $90$\% success rates in most scenarios, with convergence times nearly halved compared to baselines. Overall, this study establishes a geometry-aware, real-time approach for acoustic threat localization and advances autonomous surveillance capabilities of underwater infrastructure.

\end{abstract}

%% file: src/01_Intro.tex
\section{Introduction}
The rapid growth of cloud computing, artificial intelligence (AI), and data-driven technologies has dramatically increased the global demand for data storage and processing capacity~\cite{davenport2024ai,masanet2020recalibrating}. As traditional land-based data centers struggle with escalating power consumption~\cite{sean2024,koot2021usage}, cooling costs~\cite{katal2023energy}, land logistics~\cite{monserrate2022cloud}, and environmental impact~\cite{ewim2023impact}, the industry is increasingly exploring offshore solutions, including floating~\cite{networkocean_misc} and underwater data centers (UDCs)~\cite{microsoftunderwaterdatacenterarticle_misc} to address scalability constraints. Offshore data centers benefit from renewable energy, the ocean's cooling capability, and physical isolation from tampering, while also being protected from air-based corrosion and consequently requiring less maintenance than traditional land-based facilities~\cite{microsoft2_misc,simon2018project,hu2022packing,abner2024underwater}. Despite these advantages, UDCs introduce new physical-layer vulnerabilities arising from their unique environment. The pressure vessel (pod), housing the servers, is filled with Nitrogen gas as it is over eight times less corrosive than air; however, Nitrogen medium and the surrounding waterbody are denser than air and conduct sound much faster and further, which makes UDCs more susceptible to acoustic attacks. In particular, hard disk drives (HDDs), which constitute around 90\% of the backend storage of current cloud infrastructures along with SSDs used as fast low-capacity caches~\cite{seagate,forbes_seagate_hdd}, %\textcolor{red}{please cite the same references I wrote in the background section here}, 
are vulnerable to acoustic injections, which can lead to I/O errors, throughput degradation, and even permanent system crashes. Our prior work has shown that acoustic injections more than $6$\,meters from UDC pods pose a serious risk to data reliability and availability~\cite{sheldon2023deep,sheldon2024aquasonic}.  %\textcolor{red}{try to avoid the use of dashes as they are a typical mark for AI generated text. I already changed a few}

\begin{figure*}[t]
  \centering
  \includegraphics[width=\linewidth]{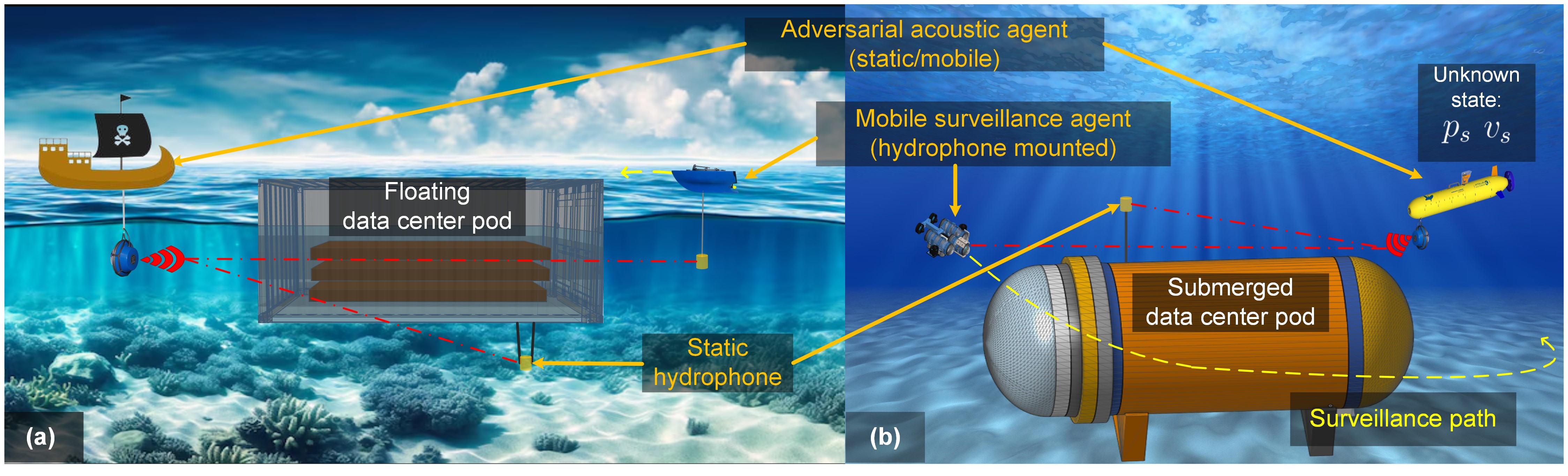}%
  \vspace{-1mm}
  \caption{An overview of our adversarial acoustic source localization (\textbf{AASL}) framework is shown for two representative configurations: (a) {floating data center pods}, and (b) {submerged data center pods}. We utilize a heterogeneous receiver pair comprising one static hydrophone mounted on the structure and one mobile hydrophone mounted on a mobile robot. The objective is to estimate and track the 3D position ($\mathbf{p_s}$) and velocity ($\mathbf{v_s}$) of an adversarial agent by integrating the proposed LC-MAP scheme that leverages acoustic geometry and locus consistency for robust localization of active targets.}
  \vspace{-4mm}
  \label{fig:problem_scenario}
\end{figure*}

Detecting and localizing the origin of acoustic attacks is vital for data center surveillance and defense. Our recent work on internal detection~\cite{blow2025detection} leveraged anomalous Position Error Signals (PES) extracted from the storage systems to identify the presence of acoustic interference. We observed that the relative performance degradation among drives offers a coarse directional cue: drives nearer to the attacker experience greater disruption. However, such inference provides only approximate directionality without revealing the absolute position or motion of the source. In this work, we advance beyond detection to achieve comprehensive 3D localization and tracking of an adversarial acoustic source.

While optical localization methods are effective for terrestrial surveillance systems, underwater vision systems are limited to a few meters under clear water and degrade rapidly in turbid or low-light conditions~\cite{islam2024computer}. Imaging sonars can extend visibility to tens of meters, but their high cost, limited refresh rate, and low angular resolution make them unsuitable for agile threat response in dynamic conditions~\cite{gaspar2023feature,munnelly2024applications}. Consequently, we explore sound source localization (\textbf{SSL}), a method widely adopted in underwater sensor networks (UWSNs) and subsea navigation~\cite{sathish2023review,li2023long,tollefsen2016source}. 

\rebuttal{However, monitoring a large data center facility against adversarial acoustic sources differs from conventional SSL in several ways. First, hydrophone arrays are difficult to scale across large, multi-pod facilities due to synchronization, wiring, calibration, and maintenance overhead~\cite{kazzazi2025clear}. Second, the adversarial source is narrowband and non-cooperative, which makes relative delay estimation ambiguous and more challenging than in cooperative broadband or pulsed localization systems. The source may maneuver within a few meters of the structure, violating far-field or planar-wave assumptions commonly used in beamforming-based localization~\cite{klippel2016holographic,ho2007source,hendricks2019automated}. Third, a mobile surveillance robot improves spatial observation capability but introduces additional uncertainty in its position and velocity estimation. Fourth, shallow-water and near-bed deployment introduces strong surface and bottom reflections, which can bias passive difference-of-arrival acoustic measurements. Finally, the direction of arrival (DOA) or angle of arrival (AOA) methods yield only bearing information~\cite{yan2018robust,li2019direction}, not generalizable for 3D localization and motion tracking of adversarial mobile targets. These factors jointly make acoustic surveillance of UDCs a distinct SSL problem that is not addressed by existing array-based or cooperative localization methods.}
%which may suffice for coarse tasks such as surface vessel detection; however, For UDC security, actionable defense demands full 3D localization and dynamic motion tracking of the adversarial agent. 

% Yet, our scenario differs fundamentally from typical SSL applications:
% \begin{itemize}
%     \item Unknown signal characteristics: Unlike cooperative transponders emitting known pulse sequences~\cite{skarsoulis2018underwater}, the attacker's signal shape is unknown, narrowband, and possibly frequency-hopping, making delay estimation at receivers more difficult and ambiguous.
%     \item Close-range motion: The source may maneuver within tens of meters to evade detection, violating planar wavefronts or far-field assumptions~\cite{klippel2016holographic,ho2007source} commonly used in beamforming models for distant ships or marine mammal localization~\cite{hendricks2019automated,zhu2020long}.
%     \item Receiver infrastructure: Deploying large hydrophone arrays around every UDC pod is logistically and economically infeasible. Moreover, large static arrays suffer from synchronization drift, multipath echoes, and potential lack of line-of-sight~\cite{kazzazi2025clear}.
%     \item Full 3D localization requirement: Direction-of-Arrival (DOA) or Angle-of-Arrival (AOA) methods yield only bearing information~\cite{yan2018robust,li2019direction}, which may be adequate for coarse tasks such as surface vessel detection. However, for UDC security, actionable defense demands full 3D localization and dynamic motion tracking of the adversarial agent.
% \end{itemize}

To overcome these challenges, we propose an adversarial acoustic source localization (\textbf{AASL}) framework. First, we introduce a novel heterogeneous receiver architecture comprising one static hydrophone mounted on the UDC pod and one mobile hydrophone attached to an autonomous surface/underwater vehicle (ASV/AUV); see Fig.~\ref{fig:problem_scenario}. \rebuttal{It reduces fixed sensing infrastructure for large multi-pod facilities while using receiver motion to provide spatial diversity over time. The system jointly measures time-difference-of-arrival (TDOA) and frequency-difference-of-arrival (FDOA) between two receiver channels and integrates them in an unscented Kalman filter (UKF) pipeline~\cite{wan2000unscented} to estimate the 3D position and velocity of the adversarial acoustic source. 
While joint TDOA-FDOA measurement and recursive filtering concepts are adapted from existing works~\cite{chen2017tdoa}, our core novelty is the Locus-Conditioned Maximum A-Posteriori (\textbf{LC-MAP}) initialization scheme that uses early TDOA-FDOA measurements to generate a geometrically consistent and physically plausible prior for the filter. Without such informed initialization, off-the-shelf filters fail or converge slowly under nonlinear, phase-ambiguous measurement conditions. Moreover, standard TDOA estimation techniques are susceptible to multipath-corrupted, narrowband signals considered in this work. To address this, we incorporate a first-order surface-bed multipath correction model and apply phase unwrapping to maintain temporal consistency in TDOA measurements. Finally, we combine the mobile receiver state uncertainty with acoustic measurement uncertainty for stable convergence of the state estimation filter.}

\rebuttal{
We evaluate the proposed AASL framework through Monte-Carlo numerical simulations, Gazebo-ROS physics simulations, and open-water field experiments. The initialization performance is tested against three baselines: Naive, Random-Sphere, and TDOA-LS across four representative source motion models: static, constant velocity, constant acceleration, and constant turn rate. Additional ablation studies evaluate the impact of multipath-aware modeling, mobile receiver state uncertainty compensation, and scalability relative to fixed-array baselines. The results show that LC-MAP achieves the highest success rate and fastest convergence, while the multipath-aware and receiver-uncertainty-aware models improve robustness under realistic shallow-water and mobile-platform conditions.} Finally, we conduct field experimental trials for floating pod setups to demonstrate the real-world integration feasibility of the proposed system. 

% also reveals that linearized filters such as the Extended Kalman Filter (EKF)~\cite{kalman1961new} diverge under the nonlinear motion geometry of mobile source-receiver pairs. In contrast, the UKF is more accurate and stable but still suffers from slow convergence or loss of track under poor initialization or ill-posed geometry. To mitigate this, we introduce a geometry-aware initialization and tracking scheme... <to finish>
% <> adaptive initialization and motion-aware tracking scheme, where the AUV's path is actively adjusted to improve localization geometry and recover lost tracks.

% \vspace{1mm}
% \noindent
In summary, the key contributions of this paper are:
\begin{enumerate}
    \item We develop a scalable acoustic threat localization framework for offshore infrastructures by utilizing a heterogeneous two-hydrophone receiver system (one static, one mobile). This configuration enables wide-area coverage and long-term autonomous surveillance without the logistical overhead of impractically many, synchronized hydrophone arrays around large underwater structures.
    \item We propose a novel LC-MAP initialization scheme which fuses highly nonlinear TDOA-FDOA observations with geometric constraints to generate physically consistent priors for state estimation filters. Integrated into a custom UKF pipeline, LC-MAP maximizes geometric observability under noisy measurement conditions, achieving higher localization success rates and faster convergence than off-the-shelf filters with standard initialization schemes.
    \item \rebuttal{We design a joint TDOA-FDOA filtering formulation to account for practical deployment uncertainties, including shallow-water surface-bed multipath bias and mobile receiver state uncertainty. These domain-aware adaptations improve robustness when acoustic measurements are corrupted by reflected paths or navigation drift.}
    \item \rebuttal{We validate the framework through Monte-Carlo simulations across diverse attack scenarios, Gazebo-ROS physics simulations, fixed-array baseline comparison, multipath modeling, and receiver uncertainty sensitivity studies.} While existing acoustic source localization approaches are mainly evaluated through simulations,  %or offline data, 
    we conduct real-world field experimental trials for floating pod scenarios to demonstrate the robust performance and practical feasibility of our proposed framework.  
    % and  rarely evaluated in prior acoustic localization studies.
\end{enumerate}
Overall, the proposed framework enables real-time localization of acoustic threats in UDCs through continuous, active monitoring. This capability allows rapid activation of defensive countermeasures to mitigate the impact of targeted injections on UDC pods, which, as demonstrated in our prior work~\cite{sheldon2024aquasonic,sheldon2023deep,blow2025detection}, can occur within $85$\,seconds of sustained acoustic excitation.

\rebuttal{The rest of this paper is organized as follows. Sec.~\ref{sec:related_work} reviews the related works on sensing mechanisms and estimation algorithms in underwater SSL, while Sec.~\ref{sec:background} presents the background and technical preliminaries for the problem of acoustic injection attacks to UDCs and outlines the relevant assumptions. Sec.~\ref{sec:method} then presents our proposed localization framework, while Sec.~\ref{sec:eval} evaluates the method through simulations and real-world field experiments. Subsequently, Sec.~\ref{sec:discussion} discusses practical deployment challenges and limitations, and the concluding remarks are in Sec.~\ref{sec:conclusion}.}

%% file: src/02_Related.tex
\section{Related Work: Acoustic Source Localization Underwater}\label{sec:related_work}

% \textcolor{red}{Sara: it might be worth here talk briefly about the attack scenarios as it is not mentioned till section 4 which is already the evaluation}
% \Adnan{Added:} 
Although the specific problem of AASL has not been addressed in prior literature, similar physical configurations and solutions have been studied in the broader domain of underwater SSL~\cite{shuanglin2025deep,desai2022review}. \rebuttal{These prior works provide useful foundations for estimating multi-channel time- and frequency-difference measurements from acoustic signals and for tracking source states using recursive filters. However, existing methods assume cooperative sources, broadband or pulsed signals, far-field propagation, or fixed receiver arrays. In contrast, adversarial acoustic attacks are non-cooperative, narrowband, and close-range, and require timely localization to enable subsequent defensive measures. These differences highlight three key research gaps:
\textbf{(i)} existing TDOA-FDOA methods are not designed for phase-ambiguous narrowband signals; \textbf{(ii)} most 3D localization systems require multi-element static arrays that are difficult to scale across distributed offshore infrastructures; and \textbf{(iii)} standard state estimation filters are sensitive to initialization and often fail to converge when no prior information of the source is available during initialization.}
% The attack scenarios considered in our work can be mapped to classical transmitter-receiver setups. For instance, a stationary attacker resembles a static transmitter-mobile receiver configuration, commonly seen in UWSNs~\cite{skarsoulis2018underwater}. In contrast, a mobile attacker maneuvering around the data center can be compared to marine mammals or AUVs, where the transmitter is dynamic, and its trajectory is not known a priori~\cite{hendricks2019automated}. 

% These analogies enable the formulation of adversarial localization problems within the context of SSL research, while introducing new challenges such as close-range motion, unknown signal structure, and the crucial need for rapid operation.

\begin{table}[t]
\centering
\caption{Three categories of underwater acoustic localization scenarios are listed with state-of-the-art solution methods.}
\small
\begin{tabularx}{\linewidth}{c|c||X}
\hline
\textbf{Transmitter} & \textbf{Receiver} & \textbf{Selected References} \\
\textbf{State} & \textbf{State} & \textbf{} \\
\hline
Mobile & Static &
TOA: \cite{chen2025underwater,liu2024litm},
AOA: \cite{wang2017unified,yan2018robust,li2019direction},
TDOA: \cite{mandic2019underwater,poursheikhali2015tdoa,liu2023efficient,valente2016real,hao2020enhanced},
RSS: \cite{chang2018target,xu2016rss,poursheikhali2021source},
TOA+AOA: \cite{kang2023efficient,jiang2021improved},
TOA+FOA: \cite{qin2024underwater},
TDOA+FDOA: \cite{chen2017tdoa,jiang2020underwater,ho2007source} \\
\hline
Static & Mobile &
TDOA: \cite{sun2023low,skarsoulis2018underwater},
AOA: \cite{gadre2008cooperative,zhang2024improved} \\
\hline
Mobile & Mobile &
TDOA+FDOA: \cite{zhang2018underwater,su2022algorithm},
TDOA+AOA: \cite{ho2023integrating} \\
\hline
\end{tabularx}
\vspace{-3mm}
\label{tab:related_work}
\end{table}

\begin{figure*}[t]
    \centering
    \includegraphics[width=\linewidth]{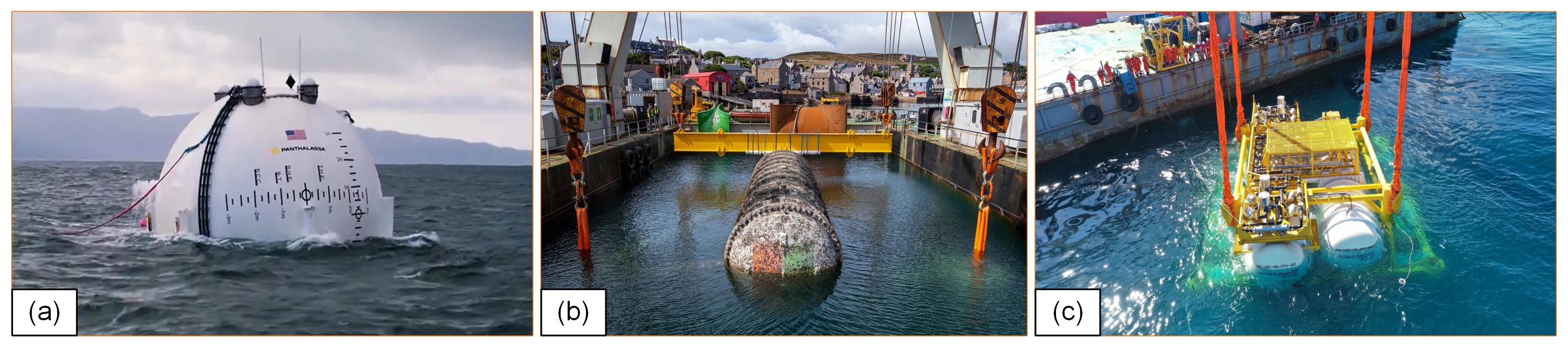}%
    \vspace{-2mm}
    \caption{\rebuttal{Three pod examples are shown for offshore industrial data centers: (a) Panthalassa's floating pod~\cite{panthalassa}; (b) Microsoft Natick's single underwater pod~\cite{microsoftunderwaterdatacenterarticle_misc}; and (c) Hailanyun's pair of underwater pods deployed at depths~\cite{hailanyun}.}}
    \label{fig:pod_setups}
    \vspace{-2mm}
\end{figure*}

\subsection{Transmitter-Receiver Setups}
Underwater SSL systems are broadly categorized into two settings: a static transmitter and mobile receiver array, or a mobile transmitter and static receiver array~\cite{erol2011survey,liu2012doubly,ferreira2023single}. Classical long-baseline systems belong to the former, where a fixed beacon on the seafloor emits known signals that are detected by a towed or autonomous hydrophone array~\cite{li2023long,chi2020design,zhu2020long}. In contrast, marine bio-acoustics represents the latter, where vocalizing animals act as mobile sources while fixed or drifting hydrophones serve as receivers~\cite{gedamke2001localization,hendricks2019automated}. These approaches are suitable for far-field scenarios (up to hundreds of kilometers) and rely on multiple receiver elements to resolve both the direction and range of the source. Recent works have explored optimal array geometries~\cite{tollefsen2016source,kim2025optimal} for accurate localization of floating/submerged acoustic sources. For instance, Baron \etal~\cite{baron2020hydrophone} demonstrate that an array of $21$ hydrophones configured in a tip-down conical pattern provides maximum spatial coverage within $350$\,Hz to $7$\,kHz frequency band for long-range source tracking. While effective, such designs are logistically complex, suffer from clock drift across receiver channels, and are often not suitable for deployment in constrained underwater environments. CLEAR~\cite{kazzazi2025clear} presents a minimal receiver architecture consisting of $N+1$ hydrophones to localize a source in $N$-dimensional space. \rebuttal{Our work explores a different design: one fixed receiver on the infrastructure and one mobile receiver on a surveillance robot. This heterogeneous configuration reduces the number of fixed sensors while using receiver motion to increase spatial diversity over time.}

% Our proposed work further reduces the number of receivers by utilizing only two hydrophones in a heterogeneous configuration: one fixed hydrophone mounted on the infrastructure, and one mobile hydrophone carried by a surveillance robot.

% Fixed beacon on seafloor + towed hydrophone array by surface vessel or submersible vehicle: known signal characteristics (\eg pulse). Marine animal (transmitter) + fixed/towed hydrophone array receiver: far-field 

% \subsection{Sound Source Localization Methods}

\subsection{Localization Approaches}
Various signal processing techniques have been proposed for underwater SSL that address environmental challenges such as unknown propagation speed~\cite{zhang2018underwater}, receiver location uncertainty~\cite{ma2023sensor,zhang2023passive}, Doppler shifts~\cite{gong2020auv}, etc. Classic approaches estimate Angle-of-Arrival (AOA) using arrays of (directional) hydrophones, but their accuracy depends on array aperture and the measurement does not reveal source range~\cite{wang2017unified,yan2018robust,li2019direction}. For complete 3D localization, the Time-of-Arrival (TOA), \ie, absolute propagation delay between the source and receivers, is measured, but this approach requires precise time synchronization~\cite{cheung2004least,shen2012accurate}. TDOA approaches rely only on relative delays between receivers, which reduces synchronization burden but still requires 3D arrays for accuracy~\cite{mandic2019underwater}. FDOA methods exploit Doppler shifts induced by relative motion between the source and receivers to infer range-rate differences~\cite{su2022algorithm}. Other approaches include Received Signal Strength (RSS)~\cite{xu2016rss,ln2019efficient} and Differential RSS (DRSS)~\cite{hu2017robust}, which utilize signal attenuation models to estimate range, but suffer from varying path loss in underwater acoustic channels and require frequent calibration.

Given the unique environmental challenges, researchers have fused multiple approaches for robust localization~\cite{kang2023efficient,qin2024underwater}. Joint TDOA-FDOA is a suitable choice for mobile-source mobile-receiver settings, where range difference (from TDOA) and range-rate difference (from FDOA) are combined into a single non-linear model. Recursive filtering solutions such as the extended Kalman filter (EKF)~\cite{kalman1961new} and unscented Kalman filter (UKF)~\cite{wan2000unscented} have been explored to handle such non-linear dynamic scenarios. However, their performance strongly depends on the initial estimate; when the source position and velocity are completely unknown (e.g., adversarial emission scenarios), the filters converge slowly or fail to track. Therefore, we propose a novel initialization scheme, LC-MAP, that provides acoustically informed priors for the filter, enabling robust convergence and tracking compared to off-the-shelf state estimation filters.

%% file: src/03_Background.tex
\section{\rebuttal{Background: Adversarial Acoustic Attack in UDCs}}\label{sec:background}

\begin{table*}[t]
\centering
\caption{\rebuttal{A qualitative comparison of sensing modalities and localization techniques for acoustic threat surveillance in large offshore infrastructures is presented. Our approach is well-suited since it minimizes sensing infrastructure and power requirements, while utilizing spatial diversity from a mobile receiver to estimate the dynamic source state. %\JI{Do I still need to worry about your Table and image captions? I spend 10 minutes on it, and it still seems like writen by an UG student.}
}}
\label{tab:method_suitability}
\small
\rebuttal{
\begin{tabular}{
>{\centering\arraybackslash}p{0.015\textwidth}|
>{\raggedright\arraybackslash}p{0.13\textwidth}|
>{\raggedright\arraybackslash}p{0.32\textwidth}|
>{\raggedright\arraybackslash}p{0.41\textwidth}
}
\hline
 \multicolumn{2}{c|}{\textbf{Category}}  & \textbf{Strengths} & \textbf{Limitations} \\
\hline
\multirow{2}{*}{\rotatebox[origin=r]{90}{Optical}}
& Visual camera
& Mid-range object detection/tracking in clear water 
& Visibility $\sim20$\,m in clear water but $\leq5$\,m in turbid water; degrades more under low light, occlusion \\
\cline{2-4}
& Underwater LiDAR
& High-resolution short-range 3D mapping
& Limited by laser scattering and turbidity; higher power demand ($75$\,mW) than passive acoustics \\
\hline
\rotatebox[origin=r]{90}{Sonar}
& Imaging sonar
& Object detection in poor visibility and mapping beyond optical range
& Lower angular resolution and refresh rate ( $\sim2^\circ$ beamwidth; $3$-$30$\,s per $360^\circ$ scan); detects 2D location rather than 3D position and velocity \\
\hline
\hline
\multirow{5}{*}{\rotatebox[origin=r]{90}{Passive acoustics}}
% & Large static hydrophone arrays
% & Accurate acoustic localization with high spatial diversity
% & High sensor count, synchronization, calibration, wiring, and maintenance burden across large or multi-pod facilities \\
% \cline{2-4}
& Hydrophone arrays
& Local acoustic monitoring on one pod
& Limited aperture; require $6N$-$8N$ receivers for $N$-pod facility; complex wiring/synchronization \\
\cline{2-4}
& GCC-PHAT for TDOA
& Robust delay estimation in broadband signals
& Does not resolve periodic phase ambiguity for narrowband/tonal attack signals \\
\cline{2-4}
& Deep learning
& Models complex propagation and environmental effects
& Requires large labeled data with various source states, multipath profiles, and attack frequencies \\
\cline{2-4}
& \textbf{Ours}
& {Minimal infrastructure, mobile spatial diversity, stable initialization under narrowband phase ambiguity}
& {Affected by mobile receiver state uncertainty and high-order multipath interference (see Sec.~\ref{subsec:acoustic_measure} and~\ref{subsec:localization_alg})}  \\
\hline
\end{tabular}
}
\end{table*}

\rebuttal{Underwater data centers introduce a new physical security vulnerability in which storage devices enclosed inside offshore or submerged pods are exposed to remote acoustic emissions. Unlike terrestrial facilities, UDC pods are surrounded by water and filled with nitrogen gas; both media transmit sound more efficiently than air, making acoustic energy coupling into the enclosure even stronger~\cite{blow2025detection}. In this setting, a remote adversary can emit narrowband acoustic signals that induce mechanical vibrations in storage devices, leading to I/O failures and service denial. This section establishes the background for the proposed localization framework by describing the acoustic injection mechanism, the adversarial source model, and the sensing assumptions that motivate passive acoustic surveillance.}
% Underwater data centers create a distinct physical attack surface because computing and storage hardware are enclosed inside offshore or submerged pods that can transmit acoustic energy from the surrounding water to internal server components. In this setting, an adversary does not need to interfere with fiber-optic communication links; instead, a narrowband acoustic source can exploit mechanical resonances of HDD-based storage and cause I/O failures or service disruption. This section establishes the background for the proposed localization framework by describing the acoustic injection mechanism, the UDC pod setting, the adversarial source model, and the sensing assumptions that motivate passive acoustic surveillance. 
%\JI{are you serious with one sentence paragraph?}

\subsection{Acoustic Injection Attacks}
\rebuttal{We consider acoustic injection attacks against remote UDC infrastructure that contains mechanically sensitive storage devices. 
% The attack does not target data communication links such as fiber-optic cables; rather, it exploits the mechanical response of HDD components inside servers. 
Acoustic injection attacks explored in literature refer to the process of emitting controlled sound waves, typically tonal or narrowband signals, from a speaker or long-range acoustic device (LRAD) toward a target computing device or electronic component to affect their functioning~\cite{shahrad2018acoustic}. Recent research~\cite{sheldon2023deep, sheldon2024aquasonic} has discovered how these attacks can be used to induce Denial-of-Service (DoS) of storage systems in servers enclosed in metal structures underwater. When the emitted acoustic wave interacts with the mechanical structure at their resonance frequencies, they induce strong vibrations that propagate through the surrounding structure, reaching the server's internal storage~\cite{bolton2018blue}. The disturbances affect the spinning platters and the actuator arm carrying the read/write heads in HDDs, causing I/O operation failure (e.g., not able to write data in the disks), data corruption (e.g., physical damage which cause not readable data in the disk), and system unavailability (e.g., application crashes). Rapid response to such acoustic exposure is essential as it may cause total system failure within approximately $85$\,s, while automated server fault-tolerance mechanisms may take up to $15$\,minutes to trigger, which is too late to prevent permanent disk damage and application crashes.} 

\rebuttal{While SSD-based systems theoretically are not susceptible to this vulnerability because of the absence of moving parts, data centers predominantly rely on backend HDD storage based on a combination of factors, including lower price, high capacity, better reliability, and data retention compared to SSDs. For instance, the vast majority of cloud workloads, around $80$-$90$\% of exabytes in cloud data centers, are stored on HDDs, and only the remaining $10$-$20$\% are stored on SSDs for fast access~\cite{seagate,forbes_seagate_hdd}. Therefore, this work focuses on developing a robust and fast localization framework capable of identifying adversarial acoustic sources within a narrow time window to allow fast recovery and prevent DoS.} 

%\textcolor{red}{please cite \url{https://www.seagate.com/blog/why-hdds-dominate-hyperscale-cloud-architecture/} and this \url{https://www.forbes.com/sites/tomcoughlin/2026/03/03/seagate-qualifies-40tb-hamr-hdds-at-two-leading-data-center-companies/}}.

\rebuttal{To validate the framework, we consider a representative UDC pod setting whose enclosure material and internal HDD rack layout are designed to resemble recent industrial UDCs such as Microsoft Natick~\cite{microsoftunderwaterdatacenterarticle_misc}.
The term ``pod'' refers to a pressure-sealed underwater enclosure that houses computing and storage hardware. Note that other industrial facilities may have different pod designs as shown in Fig.~\ref{fig:pod_setups}, and their exact acoustic vulnerability depends on material properties, internal server rack configuration, and acoustic propagation characteristics from water to the enclosure and internal storage. 
%Accordingly, we do not assume that a single attack scenario generalizes to every possible pod design.
% In this setting, once a nearby acoustic source is suspected, the surveillance system must estimate its position and motion quickly enough to support timely mitigation. 
}

\rebuttal{We model the adversarial source as an active underwater or surface agent emitting narrowband acoustic energy toward the target pod. The source may remain stationary to concentrate energy on a spot, or it may move around to search for vulnerable regions. Unlike cooperative underwater beacons, the attack signal does not have a known pulse sequence or synchronization pattern. It is a narrowband/tonal signal whose delay and Doppler signatures must be inferred from the received signal alone. This makes the localization more challenging since narrowband signals introduce phase wrapping and ambiguous TDOA estimates.
}

\subsection{Sensing Scope}
\rebuttal{We focus on passive acoustic sensing as a complementary surveillance capability rather than a replacement for camera or laser-based monitoring. Such sensors may be useful in clear water for identifying a nearby adversary. However, our goal is not general-purpose object detection, but rather an \textit{always-on surveillance} capability for UDC facilities. For such long-duration monitoring, passive acoustics is ideal since it directly senses the attack signal, operates at low power, and does not require active illumination or continuous high-bandwidth imaging. In contrast, underwater LiDARs consume about $40\times$ more power ($75$\,mW for a $15$\,m range laser)~\cite{voyis_laser} than passive acoustic sensors ($1.75$\,mW)~\cite{h2dm_hydrophone_misc}. Additionally, practical visibility for underwater cameras can degrade to only a few meters or less in seawater~\cite{yoshida2019horizontal}. They are also susceptible to turbidity, bio-fouling, and line-of-sight constraints. Moreover, compact hydrophone arrays are a viable alternative for small deployments, but they increase sensor count, wiring complexity, synchronization, and maintenance requirements in multi-pod facilities. In contrast, the proposed heterogeneous configuration seeks to minimize fixed infrastructure by using one anchor receiver on the facility and one mobile receiver carried by a surveillance robot. Table~\ref{tab:method_suitability} presents a qualitative comparison of these alternative sensing modalities and localization techniques for our target use case.
}

%% file: src/04_Method.tex
\section{Problem Formulation and Methodology}
\label{sec:method}
\begin{figure*}[t]
    \centering
    \begin{subfigure}[t]{1.05\columnwidth}
        \centering
        \includegraphics[width=\linewidth]{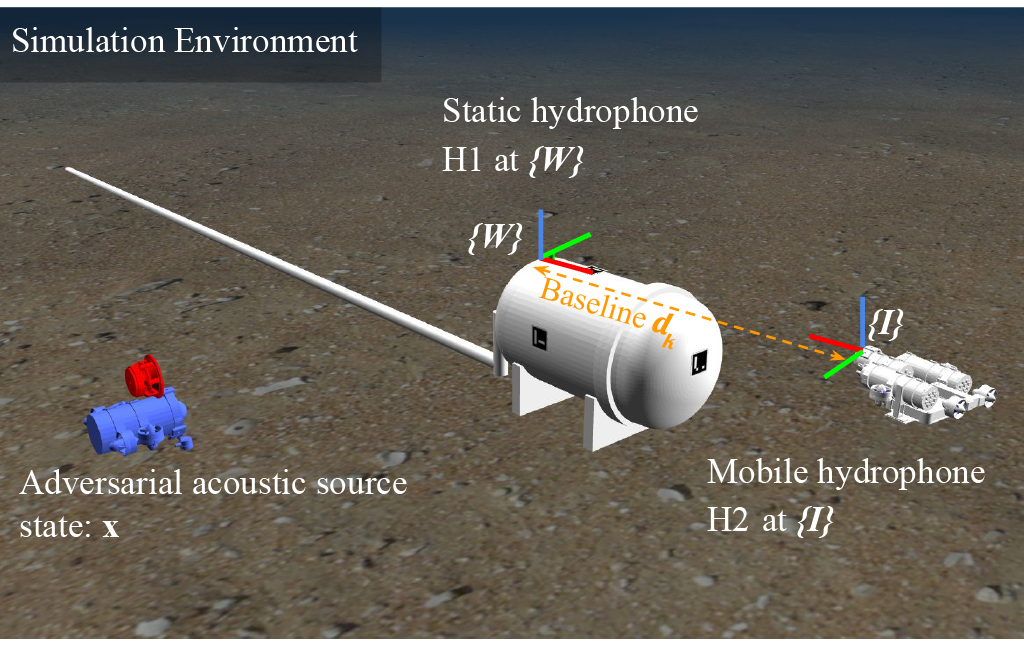}%
        \vspace{-1mm}
        \caption{Reference coordinate frames for our acoustic source localization framework: the two hydrophones H1 and H2, separated by a baseline distance $d_k$, coincide with the world frame $\{W\}$ and the AUV's inertial frame $\{I\}$, respectively. The goal is to estimate the unknown source state $\mathbf{x}$.}
        \label{subfig:ref_frames}
    \end{subfigure}%
    \hfill
    \begin{subfigure}[t]{0.94\columnwidth}
        \centering
        \includegraphics[width=\linewidth]{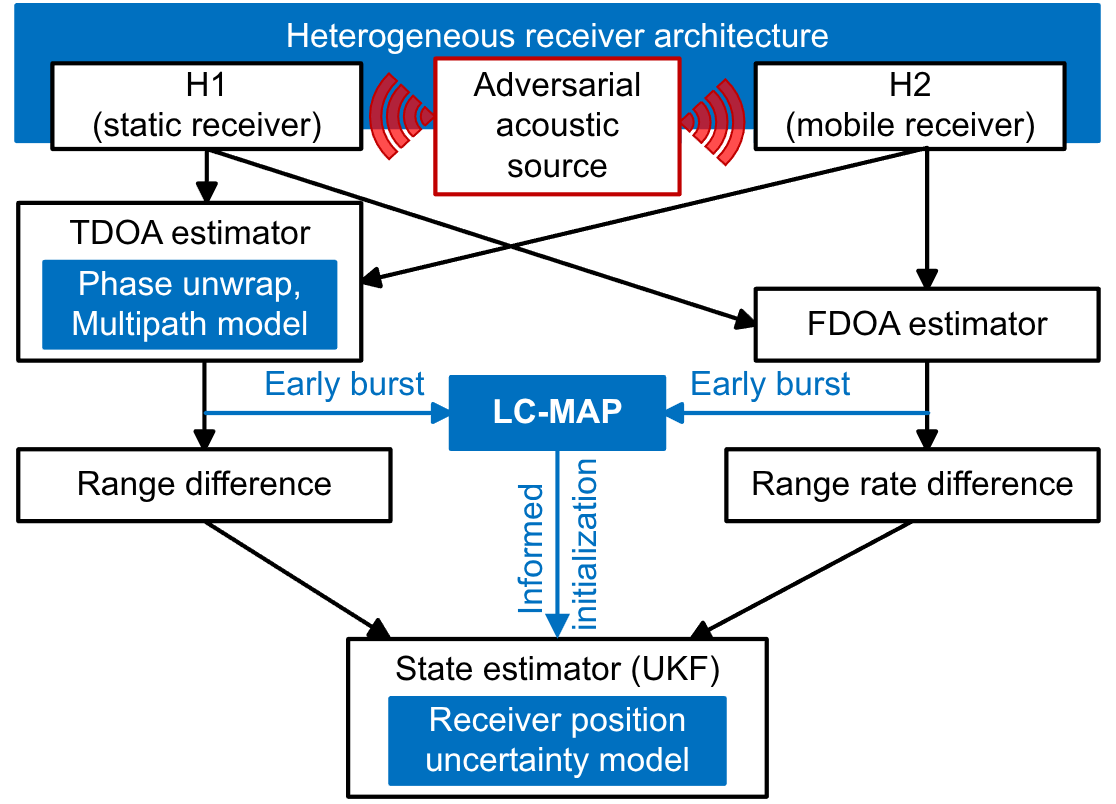}%
        \vspace{-1mm}
        \caption{\rebuttal{Our contributions in the pipeline are highlighted in blue. The TDOA estimator handles narrowband phase ambiguity and multipath interference; UKF handles the mobile receiver state uncertainty. LC-MAP utilizes early measurement bursts for a well-informed initialization.}}
        \label{subfig:system}
    \end{subfigure}
    \caption{The localization scenario and the proposed computational pipeline are illustrated. TDOA and FDOA measurements collected by one anchor hydrophone on the data center pod and one mobile hydrophone on an AUV are converted to range and range-rate differences, respectively. These measurements are jointly utilized in an Unscented Kalman Filter (UKF) to estimate the source state $\mathbf{x}$. The standard UKF diverges under uninformed initialization and nonlinear measurement dynamics, which is addressed by the proposed acoustically-informed LC-MAP module. A video demonstration can be seen here: \url{https://youtu.be/6QOY7q3n34M}.}
    \label{fig:method}
    \vspace{-3mm}
\end{figure*}

We formulate the AASL problem by combining TDOA and FDOA measurements from a two-receiver configuration (one static, one mobile), and integrating them into a UKF pipeline. \rebuttal{Our key innovation is the LC-MAP initialization scheme, which utilizes early acoustic observations and geometric consistency to yield informed priors and accelerate filter convergence. Additionally, we introduce a multipath-aware TDOA measurement model and a mobile receiver state uncertainty compensation mechanism in the UKF.} The overall system configuration and computational pipeline are illustrated in Fig.~\ref{fig:method}.

Let $\mathbf{p}_s \in \mathbb{R}^3$ and $\mathbf{v}_s \in \mathbb{R}^3$ denote the unknown position and velocity of the acoustic source, respectively. The positions and velocities of the fixed and mobile receivers are denoted by $(\mathbf{p}_f, \mathbf{v}_f)$ and $(\mathbf{p}_m, \mathbf{v}_m)$, where $\mathbf{v}_f = 0$. The speed of sound in water is $c$, and the source emits a narrowband tone of frequency $f_0$, which is assumed to be close to the data center's mechanical resonant frequency. Note that $\mathbf{p}_s$, $\mathbf{v}_s$, $\mathbf{p}_m$, and $\mathbf{v}_m$ vary over time, but for notational simplicity, the time index $k$ is omitted in the subsequent equations to represent instantaneous quantities. Therefore, at time $k$, the range from the source to the mobile hydrophone is $d_m = \lVert \mathbf{p}_s - \mathbf{p}_m \rVert$, and to the fixed hydrophone is $d_f = \lVert \mathbf{p}_s - \mathbf{p}_f \rVert$. The TDOA between the two receivers is calculated as: $TDOA = {(d_m - d_f)}/{c}$. Besides, the Doppler frequency shift, \ie, FDOA, is proportional to the relative radial velocity between the source and receivers, defined as:
% \begin{equation}
\begin{multline}
\mathrm{FDOA} = \frac{f_0}{c} \Big[
(\mathbf{v}_s - \mathbf{v}_m) \!\cdot\!
\frac{\mathbf{p}_s - \mathbf{p}_m}{\lVert \mathbf{p}_s - \mathbf{p}_m \rVert} \\
-\,
(\mathbf{v}_s - \mathbf{v}_f) \!\cdot\!
\frac{\mathbf{p}_s - \mathbf{p}_f}{\lVert \mathbf{p}_s - \mathbf{p}_f \rVert}
\Big].
\label{eqn:fdoa}
\end{multline}
% \end{equation}
By removing the scaling factors $1/c$ and $f_0/c$, these two equations are written as the range difference (RD) and range-rate difference (RRD):
\begin{multline}
RD = d_m - d_f, \\
RRD = (\mathbf{v}_s - \mathbf{v}_m) \cdot
\frac{\mathbf{p}_s - \mathbf{p}_m}{\lVert \mathbf{p}_s - \mathbf{p}_m \rVert} 
-\, (\mathbf{v}_s - \mathbf{v}_f) \cdot
\frac{\mathbf{p}_s - \mathbf{p}_f}{\lVert \mathbf{p}_s - \mathbf{p}_f \rVert}.
\label{eqn:rd_rrd}
\end{multline}
which are expressed in physical units- meters and meters per second, instead of seconds and Hertz, respectively. These geometric measurements are utilized in the subsequent state estimation process.

%\JI{after all this time, I shouldn't be commenting on Eq and symbol formatting.}

\subsection{Two-channel Acoustic Measurements}\label{subsec:acoustic_measure}
At each measurement step, the two hydrophones record time-synchronized acoustic signals of length $N$ with a sampling rate of $f_s$, from which TDOA and FDOA are estimated. The TDOA is estimated using a short-window cross-correlation. 
If $s_1[n]$ and $s_2[n]$ denote the recorded signals in the two channels, the normalized correlation over lag $\ell$ is:
\begin{equation}
R_{12}[\ell] = \frac{\sum_{n} s_1[n] , s_2[n+\ell]}{\sqrt{\sum_{n} s_1^2[n]} \, \sqrt{\sum_{n} s_2^2[n]}}.
\end{equation}
The peak lag $\ell^\star$ corresponds to the estimated TDOA: $ \Delta\hat t = {\ell^\star} / {f_s}$;
The peak lag $\ell^\star$ corresponds to the estimated TDOA: $ \Delta\hat t = {\ell^\star} / {f_s}$. \rebuttal{For underwater narrowband signals, this raw TDOA estimate suffers from phase wrapping and multipath interference, which must be addressed before it can be used reliably by the state estimation filter.}

\vspace{1mm}
\noindent
\textbf{Phase wrapping}. The estimated TDOA from cross-correlation is inherently ambiguous beyond one period for narrowband continuous signals. When the true propagation delay exceeds $\pm T_0/2$, where $T_0 = 1/f_0$, the measured delay ``wraps'' to the nearest equivalent phase, leading to discontinuities across successive frames. To maintain temporal continuity, we apply a \emph{TDOA unwrapping} step that adds or subtracts integer multiples of $T_0$ to the newly measured delay $\Delta\hat t_k$ so that it remains closest to the previous unwrapped delay $\Delta t_{k-1}$ as follows:
\begin{multline}
\Delta t_k = \Delta\hat{t}_k + n_k T_0, \\ 
n_k = \arg\min_{n \in \mathbb{Z}} 
\left| (\Delta\hat{t}_k + n T_0) - \Delta t_{k-1} \right|.
\label{eq:tdoa_unwrap}
\end{multline}

\vspace{1mm}
\noindent
\rebuttal{\textbf{Multipath interference}. In shallow-water environments, acoustic signals arrive at each hydrophone via not only the direct path but also via reflections from the water surface and the bed. These reflected arrivals corrupt the TDOA measurement with a geometry-dependent bias. To address this, we employ an image source method, where each reflector is modeled as the mirror image of the true source behind the two considered boundaries: the surface image at $p_s^{(surf)} = [x_s, y_s, -z_s]^T$ and the bed image at $p_s^{(bed)} = [x_s, y_s, -2D-z_s]^T$, where $D$ is the water depth. The multipath-corrupted range difference is modeled as:
\begin{equation}\label{eq:multipath}
    RD^{(mp)} = RD + \alpha_{surf} \cdot \delta_{surf} + \alpha_{bed} \cdot \delta_{bed}.
\end{equation}
Here, $\delta_{surf}$ and $\delta_{bed}$ are the range differences computed from the surface and bed image sources to the two receivers, respectively; $\alpha_{surf}$ and $\alpha_{bed}$ are the corresponding reflection coefficients. $\alpha_{surf}$ is negative due to the phase inversion at the water-air interface, while $\alpha_{bed}$ is positive and material-dependent.}

The FDOA measurement is derived from the Doppler shift by estimating the phase evolution of the cross-spectrum between the two channels. Let $S_1(f)$ and $S_2(f)$ denote the discrete Fourier transforms of $s_1[n]$ and $s_2[n]$. The cross-spectral density is defined as:
% \begin{equation}
$S_1(f)\,S_2^*(f) = |S(f)|\,e^{j\phi_{12}(f)}$,
% \end{equation}
where $\phi_{12}(f)$ is the cross-phase spectrum. The local phase slope yields the FDOA between the two receivers:
\begin{equation}
\Delta f_k = \frac{1}{2\pi}\,\frac{d\phi_{12}(f)}{df}.
\label{eqn:fdoa_acoustic}
\end{equation}
These front-end estimates form the measurement vector for the state estimation filter.

\subsection{Localization Algorithm: UKF-based State Estimation}\label{subsec:localization_alg}
We adapt UKF~\cite{wan2000unscented} since it captures the nonlinear relationship between the source state and the range-based measurements without requiring explicit Jacobians, unlike the EKF~\cite{kalman1961new}. Considering a constant velocity motion model, the acoustic source state is defined as:
\begin{equation}
\mathbf{x} =
\begin{bmatrix}
\mathbf{p}_s & \mathbf{v}_s & b_p & b_v
\end{bmatrix} ^T
\in \mathbb{R}^8,
\end{equation}
where $\mathbf{p}_s = [x_s ~ y_s ~ z_s]$ is the source position, $\mathbf{v}_s = [v_x ~ v_y ~ v_z]$ is the velocity; $b_p$ and $b_v$ are position and velocity biases, respectively. 
At time $k$, the two-channel received signals produce the measurements:
\begin{equation}
\mathbf{z}_k =
\begin{bmatrix}
\Delta d_k &
\Delta \dot{d}_k
\end{bmatrix}^T
=
\begin{bmatrix}
c \,\Delta t_k &
\frac{c}{f_0}\,\Delta f_k
\end{bmatrix}^T,
\end{equation}
where $\Delta t_k$ is the unwrapped TDOA, obtained from cross-correlation and unwrapping in (\ref{eq:tdoa_unwrap}), and 
$\Delta f_k$ is the FDOA, obtained 
from phase-slope estimation in (\ref{eqn:fdoa_acoustic}). 
\rebuttal{The corresponding measurement model is:
\begin{equation}
h(\mathbf{x}_k) =
\begin{bmatrix}
RD^{(mp)} & RRD
\end{bmatrix}^T,
\end{equation}
where $RRD$ is defined in~(\ref{eqn:rd_rrd}) and $RD^{(mp)}$ is the multipath-aware predicted range difference from (\ref{eq:multipath}), evaluated at the current sigma-point position. 
% Notably, this measurement prediction accounts for the same multipath bias present in $\mathbf{z}_k$, allowing the filter to correct for reflected-path corruption at every update step without requiring a separate preprocessing stage.
}

\rebuttal{
The covariance matrix $\mathbf{R}_{ac}\in\mathbb{R}^{2\times 2}$ captures the nominal acoustic TDOA-FDOA estimation noise. However, in practice, the mobile receiver position and velocity estimates may also be uncertain due to GPS drift or vehicle motion. Since the predicted TDOA-FDOA measurement directly depends on the mobile receiver state $(\mathbf{p}_m,\mathbf{v}_m)$, we account for this uncertainty by using an effective measurement covariance:
$\mathbf{R}_{eff} =
\mathbf{R}_{ac}
+
\mathbf{H}_r \mathbf{\Sigma}_r \mathbf{H}_r^\top,
$
where $\mathbf{\Sigma}_r$ is the covariance of the mobile receiver state, and $\mathbf{H}_r$ is the Jacobian of the measurement model with respect to the receiver state. In our implementation, we use a first-order isotropic approximation. Let $\mathbf{u}_m = (\mathbf{p}_s-\mathbf{p}_m) / {\|\mathbf{p}_s-\mathbf{p}_m\|}$ denote the unit vector from the mobile receiver to the source. Since $\partial RD/\partial \mathbf{p}_m=-\mathbf{u}_m$, receiver position uncertainty contributes approximately $\sigma_{\mathrm{rov}}^2$ to the $RD$ variance and velocity uncertainty contributes approximately $\sigma_{\mathrm{rov},v}^2$ to the $RRD$ variance. Thus,
\begin{equation}
\mathbf{R}_{eff}
\approx
\begin{bmatrix}
\sigma_R^2+\sigma_{\mathrm{rov}}^2 & 0\\
0 & \sigma_V^2+\sigma_{\mathrm{rov},v}^2
\end{bmatrix},
\label{eqn:reff_approx}
\end{equation}
where $\sigma_R$ and $\sigma_V$ denote the nominal acoustic $RD$ and $RRD$ noise levels, respectively.
}

Depending on the true motion behavior of the acoustic source, three motion models are considered for state estimation: constant velocity (CV), constant acceleration (CA), and constant turn rate and velocity (CTRV). Each model propagates the source state according to:
\begin{equation}
\mathbf{x}_{k+1} = f(\mathbf{x}_k) + \mathbf{w}_k, \qquad \mathbf{w}_k \sim \mathcal{N}(0,Q),
\end{equation}
where $f(.)$ represents the nonlinear motion function corresponding to the selected model and $Q$ is the process noise covariance. The standard unscented transform formulation and sigma-point weighting follow the original implementation by Wan and Van der Merwe~\cite{wan2000unscented}.
% \begin{equation}
% F =
% \begin{bmatrix}
% I_3 & \Delta t I_3 \\
% 0_3   & I_3
% \end{bmatrix}, \qquad

% Each \textit{sigma point} is mapped through the nonlinear measurement model:
% \begin{equation}
% \zeta^{(i)}_{k} = h(\chi^{(i)}_{k|k-1}).
% \end{equation}
% The predicted measurement mean and covariance are
% \begin{equation}
% \hat{\mathbf{z}}_k = \sum_i W^{(m)}_i \zeta^{(i)}_k, \qquad
% S_k = \sum_i W^{(c)}_i (\zeta^{(i)}_k - \hat{\mathbf{z}}_k)(\cdot)^T + R,
% \end{equation}
% with $R$ the measurement noise covariance.

% \vspace{1mm}
% \noindent
% \textbf{Advantages Over Existing TDOA-based Localization Schemes.}
% \begin{enumerate}
%     \item heterogeneous positioning:
%     \item Near-field applicability:
%     \item Easy synchronization: 
% \end{enumerate}

\subsection{Proposed Initialization Scheme: LC-MAP}
The proposed Locus-Conditioned Maximum A-Posteriori (LC-MAP) initialization scheme exploits the TDOA-FDOA loci and the early measurement \emph{bursts} to generate a geometrically informed prior for the UKF. The key idea is that the first TDOA measurement defines a hyperbolic locus of possible source positions relative to the two receivers, while the corresponding FDOA introduces a velocity direction along this locus. The LC-MAP scheme samples candidate positions $\mathbf{p}$ from the feasible TDOA locus and evaluates the likelihood of each candidate under the joint measurement model. For a candidate state $(\mathbf{p}, \mathbf{v})$, the measurement residual is computed as:
\begin{equation}
J(\mathbf{p}, \mathbf{v}) =
\big(\mathbf{z} - \hat{\mathbf{z}}(\mathbf{p}, \mathbf{v})\big)^{\top}
\mathbf{R}^{-1}
\big(\mathbf{z} - \hat{\mathbf{z}}(\mathbf{p}, \mathbf{v})\big),
\end{equation}
where $\mathbf{z}$ denotes the measured TDOA-FDOA vector and $\hat{\mathbf{z}}(\mathbf{p}, \mathbf{v})$ is its predicted counterpart based on the candidate state and known receiver geometry; $\mathbf{R}$ is the measurement noise covariance matrix.

\begin{table*}[t]
\centering
\small
\vspace{-1mm}
\caption{\rebuttal{Summary of the evaluation suite is presented. Each experiment has a distinct objective and thus uses appropriate test setups and metrics to quantify convergence, accuracy, robustness, scalability, etc.}}
\label{tab:eval_summary}
\rebuttal{
\begin{tabular}{
>{\centering\arraybackslash}p{0.06\textwidth}|
>{\raggedright\arraybackslash}p{0.12\textwidth}|
>{\raggedright\arraybackslash}p{0.24\textwidth}|
>{\raggedright\arraybackslash}p{0.26\textwidth}|
>{\raggedright\arraybackslash}p{0.16\textwidth}
}
\hline
\multicolumn{2}{c|}{\textbf{Evaluation type}} & \textbf{Objective} & \textbf{Implementation setup} & \textbf{Metrics} \\
\hline
\multirow{4}{*}{\rotatebox[origin=r]{90}{Monte Carlo simulation }}
& Initializer comparison
& Compare proposed LC-MAP against $3$ baseline initialization schemes
& $100$ trials for each source motion model: static, CV, CA, and CTRV
& Convergence time, success rate, RMSE \\
\cline{2-5}
& Fixed-array baseline comparison
& Compare proposed heterogeneous architecture against conventional arrays
& $N$-pod facility; $6$-sensor array and $8$-sensor array versus our static-mobile configuration
& RMSE, runtime, receiver scaling \\
\cline{2-5}
& Multipath ablation
& Evaluate the effect of shallow-water surface-bed reflections
& $100$ trials with $\alpha_{surf} = -0.9$, $\alpha_{bed} = 0.3$; direct-path-only versus multipath-aware filter
& Convergence rate, RMSE, final error, maximum error \\
\cline{2-5}
& Receiver uncertainty sensitivity
& Assess robustness to mobile hydrophone position and velocity errors
& Sweep over receiver position uncertainty $\sigma_{\mathrm{rov}}=0$-$70$\,cm, with and without compensation
& RMSE \\
\hline
Gazebo-ROS
& Physics simulation
& Validate end-to-end pipeline with robotic agents
& $60$\,m deep UDC facility, mobile receiver AUV, and adversarial acoustic source AUV
& Tracking error, RMSE \\
\hline
Field trial
& Shallow-water experiment
& Demonstrate real-world feasibility in shallow-water floating conditions
& $15$ lake trials using two ASVs, GPS telemetry, and acoustic measurements with hydrophones
& TDOA/FDOA modeling error, tracking error\\
\hline
\end{tabular}
}
\vspace{-3mm}
\end{table*}

\begin{figure*}[t]
    \centering
    \includegraphics[width=0.9\linewidth]{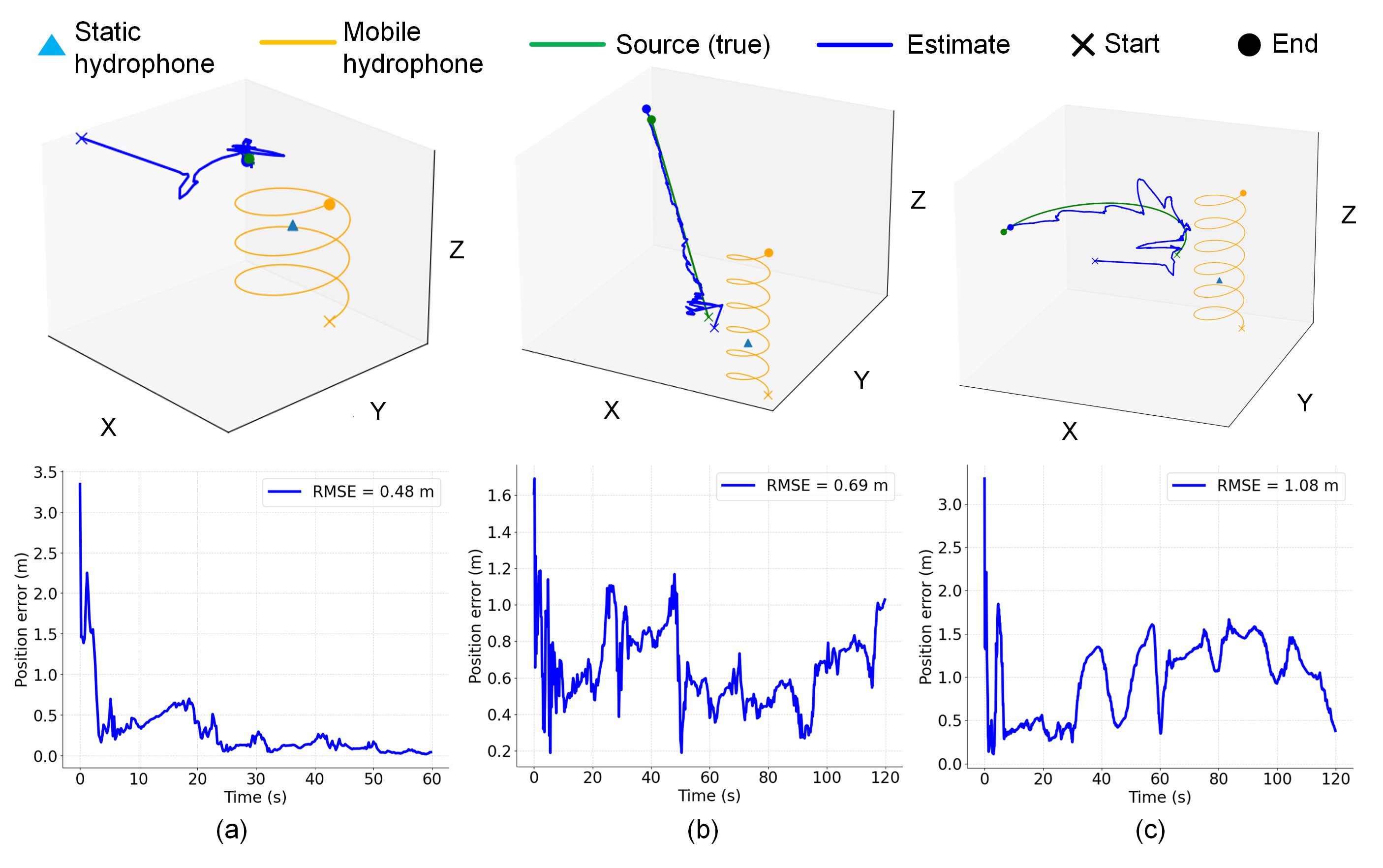}
    \vspace{-2mm}
    \caption{Three examples from Monte Carlo simulation are shown; the source behaviors are: (a) static, (b) constant velocity, and (c) constant turn rate. The static-mobile hydrophone pair collects TDOA and FDOA measurements as the latter follows a 3D helical trajectory (orange). The source path and tracking results are shown in Green and Blue, respectively. Corresponding position errors over time and their root mean square metrics are reported in the bottom row.}
    \label{fig:sim_result}
    \vspace{-4mm}
\end{figure*}

To avoid poorly conditioned configurations, a Fisher Information Matrix (FIM) term is incorporated to favor \emph{geometrically informative} regions. The ultimate initialization is obtained by maximizing: 
\begin{equation}
(\mathbf{p}_0, \mathbf{v}_0)_{\text{LC-MAP}} =
\arg\min_{\mathbf{p}, \mathbf{v}}
\Big[
J(\mathbf{p}, \mathbf{v}) - \lambda \log \det \mathbf{F}(\mathbf{p}, \mathbf{v})
\Big],
\end{equation}
where $\mathbf{F}(\mathbf{p}, \mathbf{v})$ is the local FIM and $\lambda$ controls the weight of the geometric conditioning term. Intuitively, this process prioritizes not just the most probable but also the well-conditioned initial points along the measured TDOA locus. As such, even if the initial guess is far from the true state, the high observability (\ie, larger FIM) allows the filter to adapt rapidly. To validate the benefits of LC-MAP initialization scheme, we compare its performance against three baselines:
\begin{enumerate}[label=$\bullet$]
    \item Naive: initializes the source at the geometric midpoint between the two hydrophones.
    \item Random sphere sampling (\textit{aka} Random): randomly samples candidate positions on a spherical surface consistent with the first TDOA measurement and selects the point yielding the minimum residual.
    \item TDOA-LS: iteratively refines the source position using a single TDOA measurement through least-squares updates until convergence.
\end{enumerate}

%% file: src/05_Evaluation.tex
\section{Experimental Evaluation}\label{sec:eval}
\rebuttal{The proposed localization and tracking framework is validated in three levels: (i) Monte Carlo numerical simulations, (ii) Gazebo-ROS physics simulations, and (iii) real-world shallow-water field experiments. Table~\ref{tab:eval_summary} summarizes the performed evaluations, their objectives, test settings, and adopted performance metrics. We first analyze the convergence behavior of the LC-MAP initialization scheme under different adversarial motion models. We then demonstrate the complete end-to-end localization pipeline in a physics-based underwater simulation environment. Additionally, we compare the proposed heterogeneous receiver configuration against fixed-array baselines, evaluate the effects of multipath-aware modeling, and analyze the sensitivity of mobile receiver uncertainty.
Finally, we validate the practical feasibility through open-water field trials using autonomous surface vehicles and real hydrophone measurements.}

\begin{figure*}[t]
    \centering
    \includegraphics[width=0.95\linewidth]{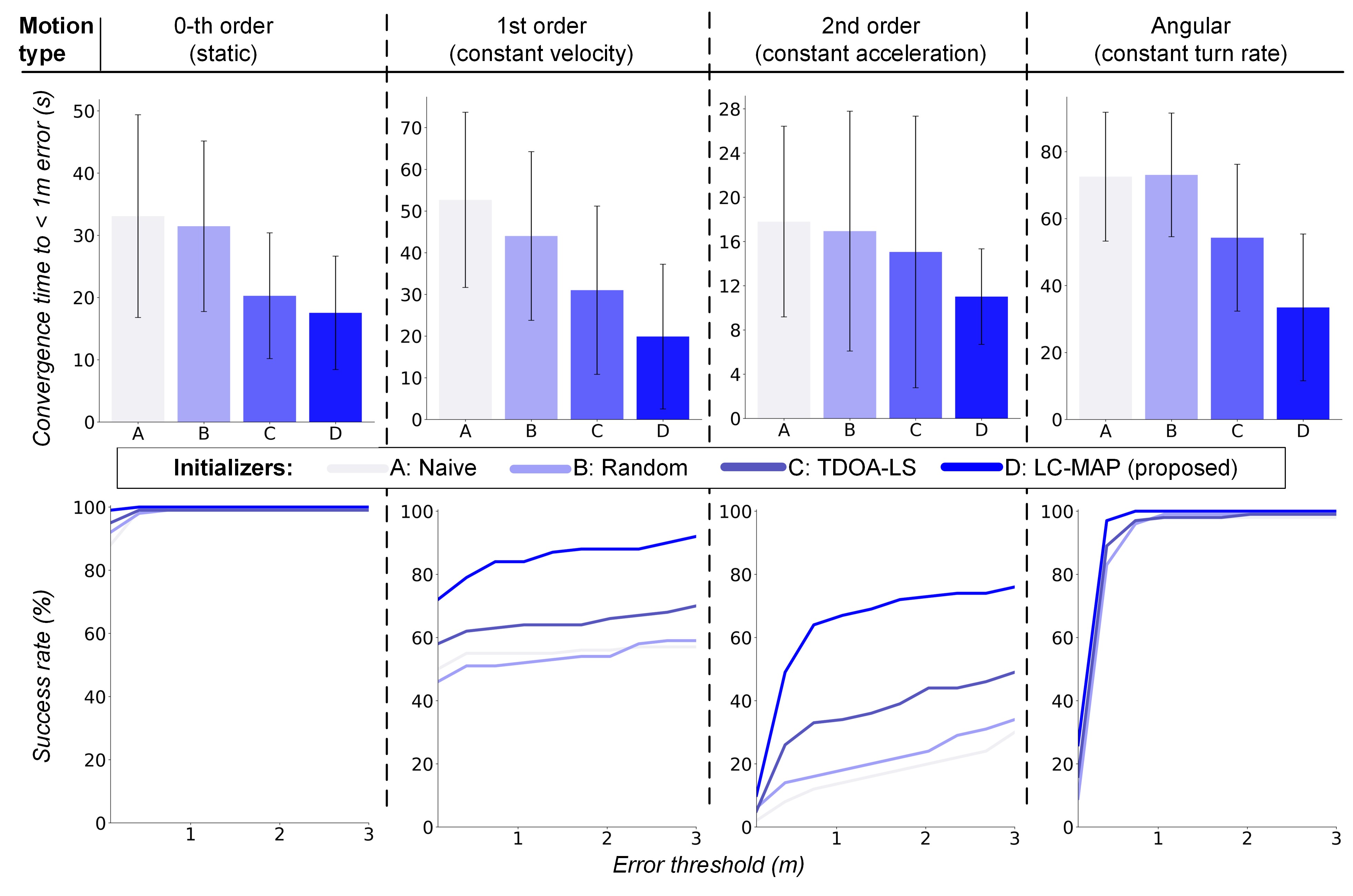}
    % \vspace{-4mm}
    \caption{Four initialization methods are evaluated across four types of source motion using $100$ Monte Carlo simulations. The top row shows the average time required for the filter to achieve a localization error below $1$\,m; the proposed LC-MAP initializer consistently converges faster than the others. The bottom row presents the success rate, defined as the proportion of trials in which the localization error falls below a given threshold within $120$\,s. While all methods succeed in the static case, LC-MAP demonstrates superior performance and reliability in dynamic scenarios.}
    \label{fig:init_comparison}
    \vspace{-3mm}
\end{figure*}

\subsection{Metrics for Initializer Performance}\label{subsec:metrics}
To assess the performance of four initialization schemes, we employ two metrics: (i) convergence time and (ii) success rate. The convergence time measures how quickly the estimate approaches the true location within a specified error threshold ($1$\,m). It reflects the quality of the initial guess and the filter's ability to quickly adapt during early iterations. The success rate quantifies the proportion of trials in which the filter converges to the true location within a specified error threshold (ranging from $0.1$\,m to $3$\,m) and within a fixed time window ($120$\,s). This metric captures the overall robustness of each initialization scheme under different motion models. 
%\JI{Missing information on attack scenarios: how many attack scenarios were considered in field trials and Gazebo simulations?}\Adnan{Added:} 
The analyses consider both static and mobile attacker scenarios. For the mobile case, we test constant-velocity and constant-acceleration motions to mimic energy-concentrated attacks, as well as circumnavigating paths to represent attackers probing for vulnerable regions.

\subsection{Monte Carlo Numerical Simulation}\label{subsec:monte_carlo}
The proposed framework is first validated through extensive numerical simulations. TDOA-FDOA measurements are recorded at a rate of $5$\,Hz and synthesized with white Gaussian noise. The static hydrophone is placed at a depth of $0.96$\,m below the origin, with the mobile hydrophone following a 3D helical trajectory around it; see Fig.~\ref{fig:sim_result}. Such a path enhances geometric observability compared to linear or planar motion patterns. Four source motion models are tested: static (zero order), constant velocity (first order), constant acceleration (second order), and constant turn rate. For each source motion and initializer type, $100$ Monte Carlo trials are conducted, with the initial source position randomly sampled within $\pm20$\,m radius from the origin.

% Figure~\ref{fig:sim_result} illustrates three representative runs for a static, constant-velocity, and constant turn rate source. 
Fig.~\ref{fig:init_comparison} summarizes the average convergence time and success rate across all four initializers. The proposed LC-MAP scheme consistently achieves higher success rates than the baseline methods, and reduces the average convergence time by nearly $50\%$ compared to the naive initializer. Note that the shorter convergence time observed for the second-order motion results from a lower overall success rate, as failed runs are excluded from the convergence time statistics.

\begin{figure*}[t]
    \centering
    \includegraphics[width=0.95\linewidth]{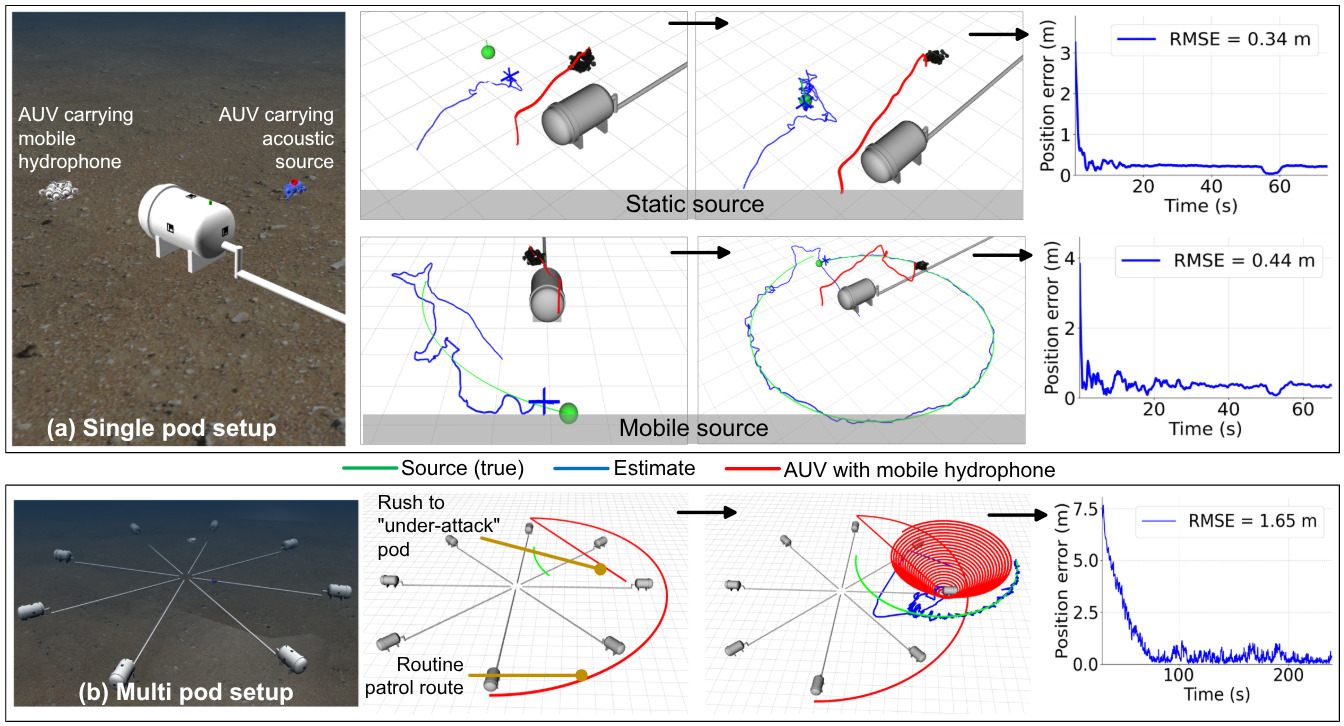}
    \caption{Representative results from the Gazebo-ROS physics simulator: (a) The UDC pod, the acoustic source, and the hydrophone receivers are rendered in a $60$\,m deep open water environment; and (b-c) Rviz snapshots for static and mobile source localization, respectively. Early and steady-state tracking results are illustrated from left to right; corresponding position error curves show sub-meter level root mean square error (RMSE) of position, indicating stable convergence over time.}
    \label{fig:gazebo_result}
    \vspace{-3mm}
\end{figure*}

\subsection{Gazebo Physics Simulation}
We further demonstrate our framework in a ROS-Gazebo~\cite{koenig2004design} physics-based simulation environment. The underwater world is adapted from the UUV Simulator~\cite{manhaes2016uuv}. We place a custom data center pod at $60$\,m depth and integrate digital replicas of two custom-built AUVs: NemeSys~\cite{abdullah2025nemesys} and CavePI~\cite{gupta2025demonstrating}. While Gazebo natively supports hydrodynamics and sensor physics \rebuttal{($1$\,ms update step)}, it lacks built-in components for acoustic propagation and hydrophone sensing. Therefore, we develop custom sensor plugins to simulate an omnidirectional acoustic source and two hydrophone receivers. The acoustic channel is modeled with an inverse-distance attenuation profile; \rebuttal{the received signal includes additive white Gaussian noise and first-order reflections from the bed and water surface}. The fixed hydrophone is mounted atop the pod, and the mobile hydrophone is attached beneath the NemeSys AUV. The CavePI AUV carries the acoustic source, acting as the adversarial agent.
% The pod is labeled with multiple fiducial AruCo markers~\cite{garrido2014automatic} for the AUV's visual pose estimation.
Once the internal detection module is triggered, the NemeSys AUV initiates a surveillance maneuver, recording TDOA and FDOA measurements while navigating around the pod. The whole framework is wrapped in ROS to synchronize the AUV's pose, two-channel hydrophone recordings, TDOA-FDOA calculation, and state update at \rebuttal{$5$\,Hz rate}. 

Fig.~\ref{fig:gazebo_result}\,a shows two experiments conducted in the Gazebo simulator for the single pod scenario. The surveillance robot executes a 3D trajectory around the pod, which enhances geometric observability in all three dimensions. The corresponding localization errors for both static and mobile source scenarios indicate that LC-MAP rapidly locks onto the true source position within the first $5$-$10$ seconds and maintains sub-meter tracking accuracy throughout the simulation.

\rebuttal{Fig.~\ref{fig:gazebo_result}\,b extends the evaluation to a multi-pod scenario. The surveillance robot initially follows a routine patrol path around the facility and, once an acoustic attack is detected, redirects toward the affected pod. After reaching near the hydrophone of the ``under-attack'' pod, the robot executes a spiral maneuver to collect geometrically diverse TDOA-FDOA measurements. The resulting trajectory and error curve show that the proposed pipeline remains stable in the larger facility-scale setup, achieving an RMSE of $1.65$\,m.}

\subsection{Comparison: Fixed Hydrophone Array}
\begin{table}[t]
\centering
\caption{\rebuttal{Fixed-array SSL methods are compared to the proposed static-mobile receiver configuration for an $N$-pod facility. Hardware requirements and performance are reported for an example facility with $N=8$ pods, as shown in Fig.~\ref{fig:gazebo_result}\,b. Rx:Receivers}}
\label{tab:array_scaling}
\small
\rebuttal{
\begin{tabularx}{\columnwidth}{l|c|c|c}
\hline
\textbf{Metric} &
\textbf{OSLS~\cite{huang2000passive}} &
\textbf{SDP~\cite{lui2009accurate}} &
\textbf{Ours} \\
\hline
Rx/pod &
$6$ &
$8$ &
$1$ + $1$ shared mobile \\
\hline
Total Rx &
$6N = 48$ &
$8N = 64$ &
$N+1 = 9$ \\
\hline
Estimate &
Position &
Position &
Position + velocity \\
\hline
RMSE &
$1.41$\,m &
$0.94$\,m &
$\mathbf{0.69}$\,m \\
\hline
Update time &
$\mathbf{0.30}$\,ms &
$8.03$\,ms &
$0.49$\,ms \\
\hline
\end{tabularx}
}
\vspace{-2mm}
\end{table}

\begin{figure*}[t]
    \centering
    \includegraphics[width=\linewidth]{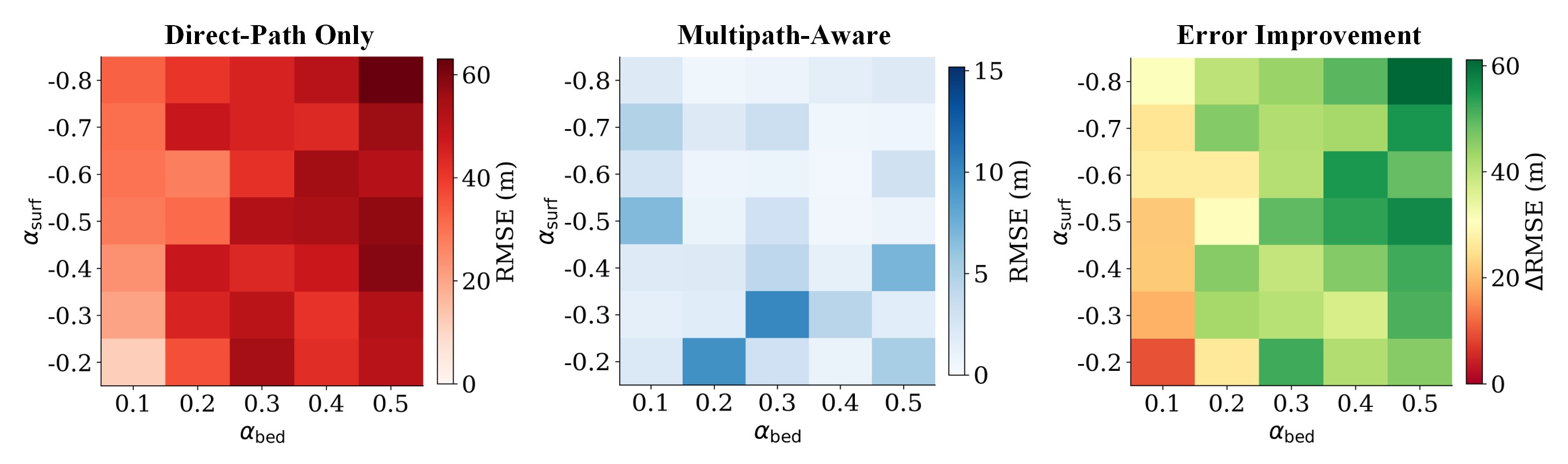}%
    \vspace{-1mm}
    \caption{\rebuttal{Sensitivity to multipath interference is analyzed by sweeping surface and bed reflection coefficients. For each pair of $\alpha_{surf}$ and $\alpha_{bed}$, $10$ Monte Carlo trials are performed with randomly sampled initial source positions. The left and middle panels show the average RMSE for the direct-path-only and multipath-aware models, respectively. The right panel shows the RMSE improvement due to incorporating the multipath-aware measurement model.}}
    \vspace{-2mm}
    \label{fig:multipath_coeff_sweep}
\end{figure*}

\rebuttal{To examine the scalability trade-off, we compare the proposed heterogeneous receiver configuration with two fixed-array baselines: the one-step least squares (OSLS)~\cite{huang2000passive} that uses $6$ receivers and the semi-definite programming (SDP) relaxation~\cite{lui2009accurate} that uses $8$ receivers. These baselines represent conventional array-based localization, where non-coplanar hydrophones provide TDOA measurements, and the source position is estimated from range difference constraints. Note that these methods do not handle the narrowband phase ambiguity problem. Therefore, we downscale their array dimensions to ensure that the receiver separation is within half-wavelength, avoiding ambiguous phase for a fair comparison. As summarized in Table~\ref{tab:array_scaling}, the OSLS method provides the fastest update rate ($0.30$\,ms) due to its closed-form formulation, but yields the largest localization error ($1.41$\,m) because each step is solved independently without temporal smoothing. The SDP method improves accuracy by solving a global optimization problem, but incurs much higher computation time ($8.03$\,ms). In contrast, the proposed method achieves the lowest RMSE ($0.69$\,m) with a moderate update rate of $0.49$\,ms by recursively fusing TDOA-FDOA measurements over time.
}

\begin{table}[t]
\centering
\caption{\rebuttal{Effect of multipath-aware modeling is analyzed over $100$ Monte Carlo trials. A trial is considered converged if position RMSE remains below $30$\,m. Even under this relaxed threshold, the direct-path-only model converges in only $9\%$ of trials, whereas the multipath-aware model remains stable in all trials. The final error indicates the position error at the end of $120$\,s long trial. All error statistics are computed over converged trials only.}}
\label{tab:multipath_ablation}
\small
\rebuttal{
\begin{tabular}{l||c|c}
\hline
\textbf{Metric} &
\textbf{Direct-path-only} &
\textbf{Multipath-aware} \\
\hline
Convergence ($\uparrow$) &
$9\%$ &
$100\%$ \\
\hline
RMSE ($\downarrow$) &
$25.71 \pm 2.44$\,m &
$0.98 \pm 2.15$\,m \\
\hline
Final error ($\downarrow$) &
$10.10 \pm 3.57$\,m &
$0.15 \pm 0.97$\,m \\
\hline
Maximum error ($\downarrow$) &
$106.16 \pm 15.83$\,m &
$9.27 \pm 22.40$\,m \\
\hline
\end{tabular}
}
\end{table}

\rebuttal{More importantly, the proposed architecture significantly reduces sensing infrastructure for multi-pod deployments; an example scenario is illustrated in Fig.~\ref{fig:gazebo_result}\,b. For an $N$-pod facility, the OSLS and SDP baselines require $6N$ and $8N$ fixed receivers, respectively, whereas the proposed system requires only $N+1$ receivers: one fixed hydrophone per pod and one shared mobile receiver. This reduces wiring, synchronization, calibration, and maintenance. Furthermore, by exploiting receiver motion, the proposed method estimates both the source position and velocity at each update step; in contrast, the fixed arrays provide only position estimates.}

\subsection{Ablation: Multipath-aware Localization}\label{subsec:multipath_ablation}
\rebuttal{To evaluate the multipath-aware measurement model, we conduct a Monte Carlo ablation study where the acoustic environment contains surface and bed reflections. We compare two variants: a direct-path-only model that ignores reflected signals, and the proposed multipath-aware model that incorporates the surface and bed reflection correction in the range difference prediction. The source's initial position is randomly sampled over the facility region for $100$ trials. A $20$\,m deep environment is considered with reflection coefficients $\alpha_{surf} = -0.9$ and $\alpha_{bed} = 0.3$. Unlike the earlier initializer comparison, which uses a strict success criterion of reaching <$1$\,m position error, this ablation focuses on filter stability. The direct-path-only filter rarely reaches the sub-meter threshold and almost always diverges. Therefore, we use a separate convergence criterion: a trial is marked as divergent if its position RMSE exceeds $30$\,m.} 

\begin{figure}[t]
\centering
% \vspace{-1mm}
\includegraphics[width=\linewidth]{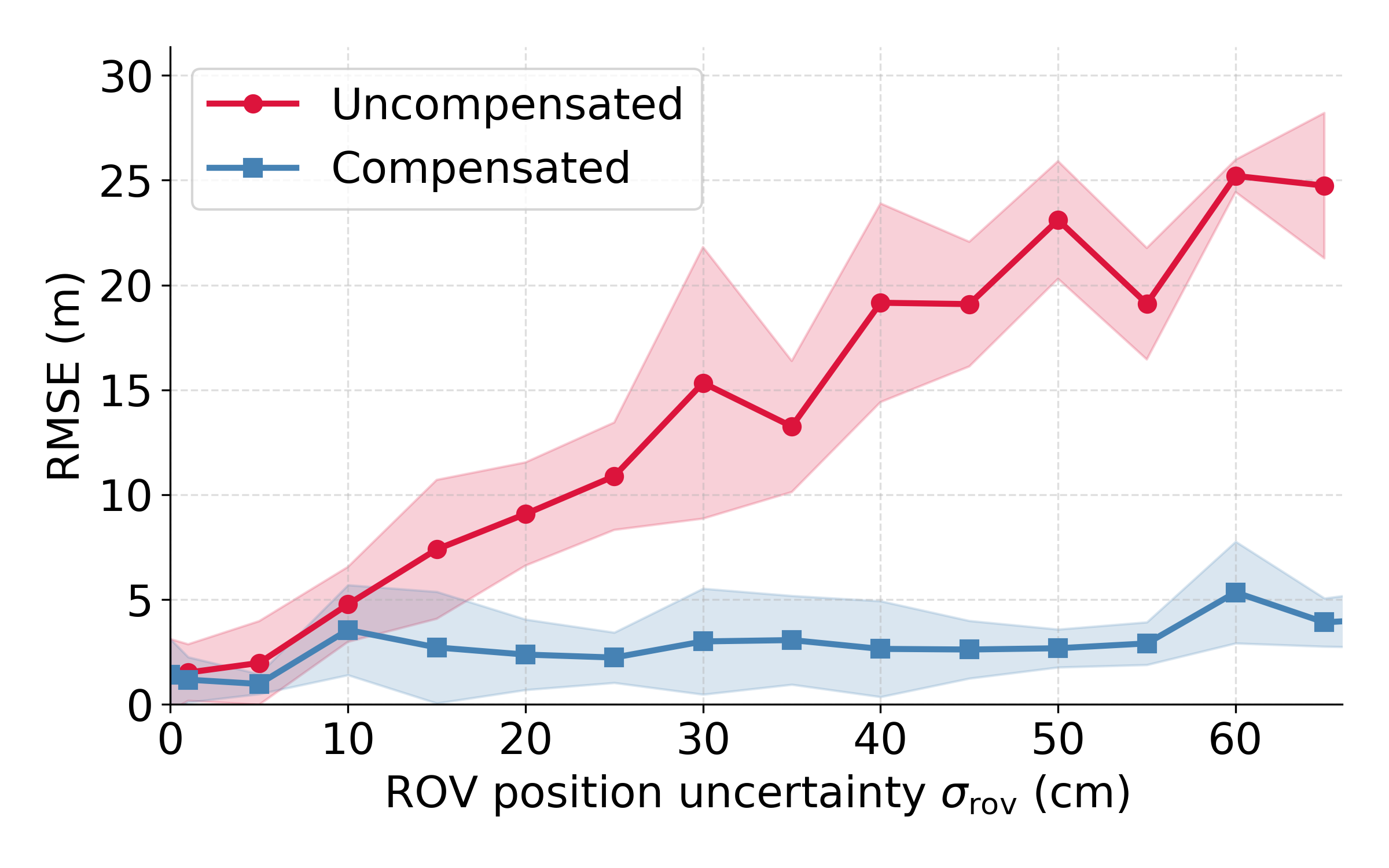}%
\vspace{-2mm}
\caption{\rebuttal{Sensitivity to mobile receiver uncertainty is analyzed in terms of localization RMSE. The uncompensated filter uses nominal acoustic measurement covariance $\mathbf{R}_{ac}$, while the compensated filter uses the effective covariance $\mathbf{R}_{eff}$ that accounts for receiver position and velocity error. Shaded regions indicate one standard deviation across $10$ Monte Carlo trials.}}%
\label{fig:rov_uncertainty}
\vspace{-3mm}
\end{figure}

\begin{figure*}[t]
    \centering
    \includegraphics[width=0.95\linewidth]{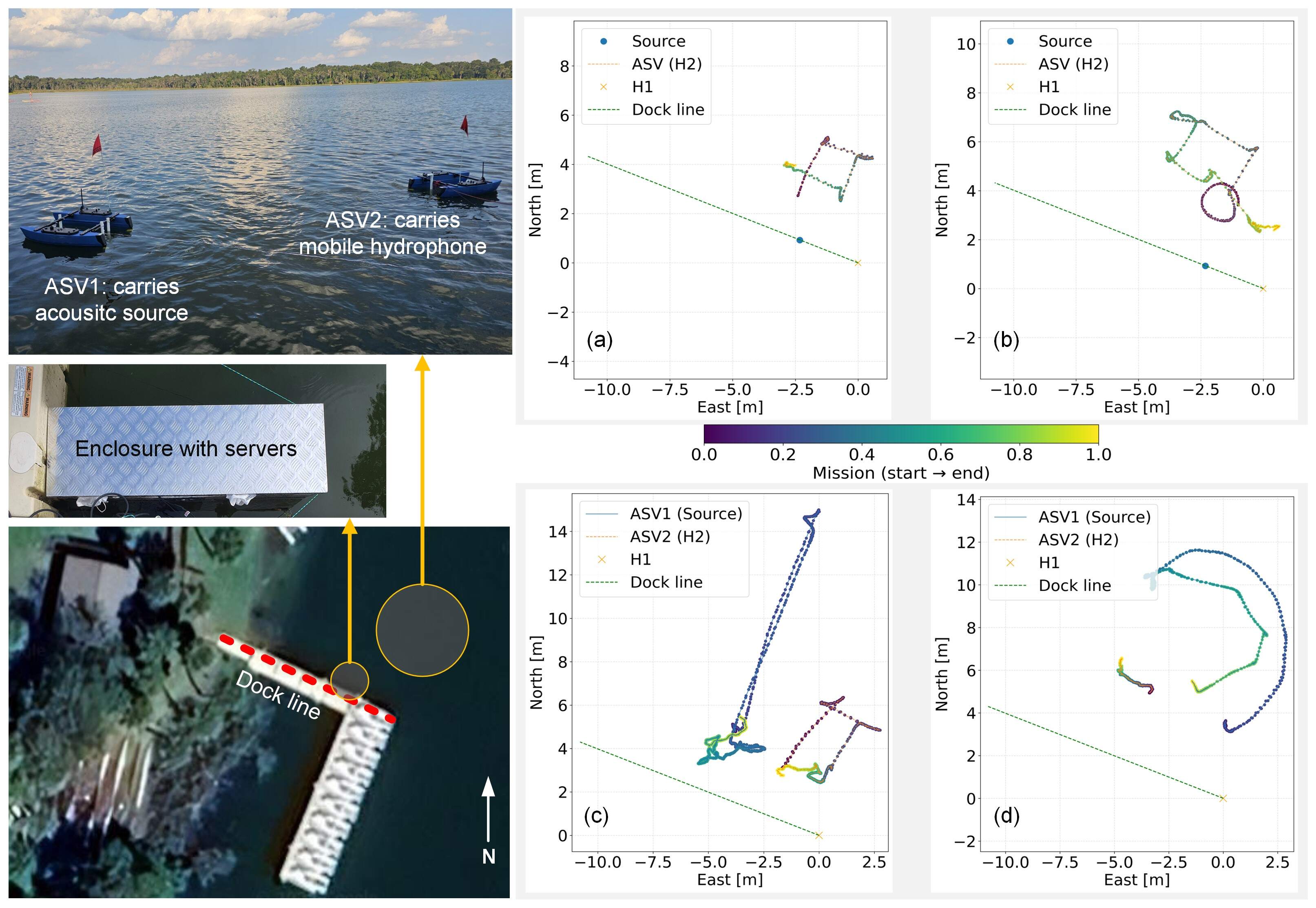}%
    \vspace{-1mm}
    \caption{The open-water test site and representative trials are shown. Two ASVs carry the acoustic source and the mobile hydrophone (H2), while the other hydrophone (H1) is suspended from the dock. The hydrophone-equipped ASV executes various patterns to collect TDOA-FDOA measurements from the source, while the source may (a,b) remain stationary, or move along (c) straight, or (d) arc-shaped paths to emulate adversarial acoustic behavior.}
    \label{fig:field_setup}
    \vspace{-4mm}
\end{figure*}

\rebuttal{As shown in Table~\ref{tab:multipath_ablation}, when multipath propagation occurs but is ignored by the TDOA measurement model, the filter converges in only $9\%$ of the trials and produces large errors even for the converged cases, with an average RMSE of $25.71$\,m. In contrast, the multipath-corrected model always converges (within $30$\,m RMSE threshold) and reduces the average RMSE to $0.98$\,m. Although the maximum error appears high, these peaks occur during the initial transient phase. The much lower final error of $0.15$\,m confirms that, after the transient period, the multipath-aware filter converges and remains stable.}

\rebuttal{To further evaluate the multipath sensitivity, we sweep the two reflection coefficients over a range of surface and bed conditions. For each coefficient pair, we run $10$ Monte Carlo trials and compute the position RMSE for both the direct-path-only and multipath-aware models.
Fig.~\ref{fig:multipath_coeff_sweep} shows that the direct-path-only model produces large errors even when the reflection is minimal ($\alpha_{surf} = -0.2$ and $\alpha_{bed} = 0.1$). In contrast, the multipath-aware model substantially reduces RMSE for all coefficient pairs. Overall, these results highlight that shallow-water multipath-reflected signals can severely bias TDOA-based localization if unmodeled, and that even a first-order surface-bed reflection model substantially improves performance.}

\subsection{Sensitivity to Mobile Receiver State Uncertainty}
\label{subsec:rov_uncertainty}

\rebuttal{We conduct this sensitivity analysis by perturbing the mobile hydrophone position and velocity during TDOA-FDOA measurement. The position uncertainty $\sigma_{\mathrm{rov}}$ is swept from $0$ to $70$\,cm, which is close to half-wavelength for a $1$\,kHz signal. The velocity uncertainty is scaled proportionally ($0.4\times\sigma_{\mathrm{rov}}$). We compare two variants: an uncompensated UKF using only the nominal acoustic measurement covariance ($\mathbf{R}_{ac}$), and a compensated UKF using the effective covariance $\mathbf{R}_{eff}$ from~(\ref{eqn:reff_approx}). For each uncertainty level, $10$ Monte Carlo trials are performed with random source initializations, and the mean RMSE is calculated.
Fig.~\ref{fig:rov_uncertainty} shows that the uncompensated filter degrades rapidly as the position error increases. In contrast, the compensated formulation remains more stable, keeping the RMSE within $3$\,m for most levels. This confirms that accounting for mobile receiver state uncertainty prevents the filter from becoming overconfident and improves robustness against position error in realistic, noisy underwater measurement conditions.
}

\begin{figure*}[t]
    \centering
    \includegraphics[width=\linewidth]{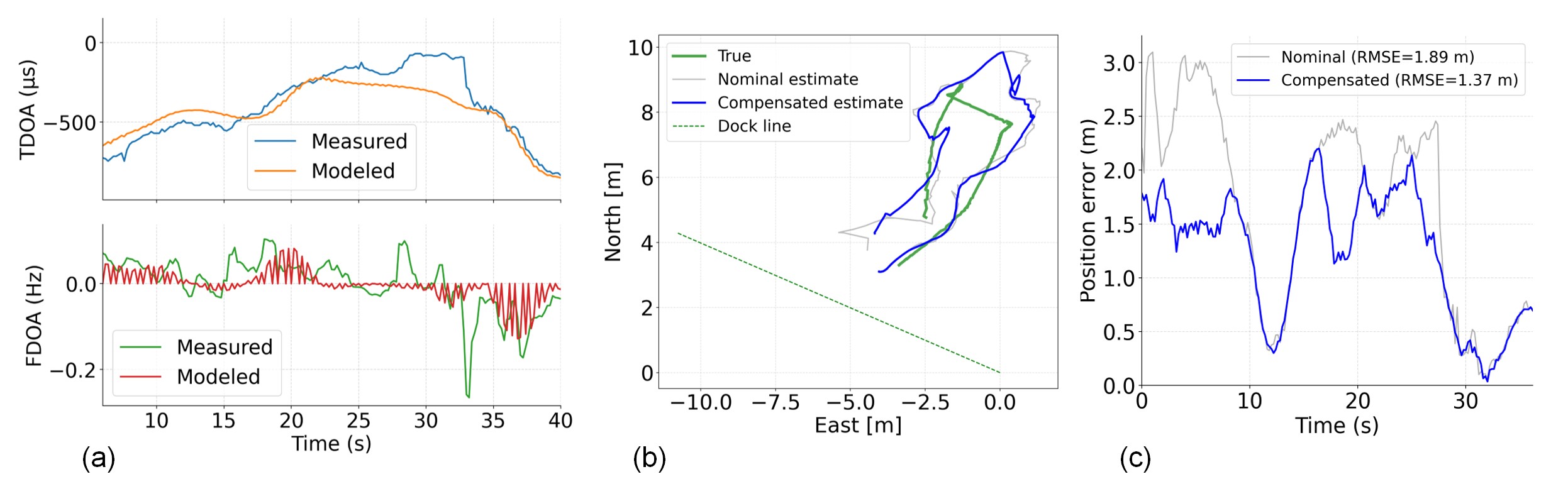}
    
    \caption{\rebuttal{Representative example from the lake trial is illustrated: (a) comparison between measured and modeled TDOA-FDOA, illustrating the acoustic noise characteristics; (b) tracking performance of the nominal (uncompensated) and the compensated filter, where the latter incorporates multipath correction and mobile receiver uncertainty compensation; (c) position error over time, showing that the compensated model improves RMSE from $1.89$\,m to $1.37$\,m under real-world conditions.}}
    \vspace{-2mm}
    \label{fig:field_result}
\end{figure*}

\subsection{Field Experiments}

Following the simulation tests, we evaluate the proposed localization system in a shallow-water lake environment. As shown in Fig.~\ref{fig:field_setup}, the acoustic source and the mobile receiver (H2) are mounted beneath two Blue Robotics BlueBoat\texttrademark{} ASVs. The static hydrophone (H1) is suspended from the dock, with its surface projection defined as the global origin. Both ASVs continuously log GPS positions, velocities, and other telemetry data, which are synchronized with the acoustic signal to evaluate localization accuracy.

\vspace{1mm}
\noindent
\textbf{Attack scenario}. A total of $15$ trials are performed to evaluate both static and dynamic adversarial attack scenarios. In the first set of trials, the acoustic source remains stationary while the mobile receiver executes simple geometric patterns (e.g., circle, square, lawnmower) to capture diverse TDOA-FDOA observations. In subsequent tests, the source is also made mobile, following straight, arc-shaped, and U-shaped trajectories. Essentially, these maneuvers mimic a stealthy adversarial agent exploring vulnerable regions around the structure. The source-carrier ASV operates within a $15$\,m radius from the anchor (\ie, static hydrophone), while the hydrophone-carrier ASV maintains a separation of $1$\,m to $6$\,m from the anchor to avoid collision. The static and mobile hydrophones are suspended at depths of $0.91$\,m and $1.19$\,m, respectively, while the acoustic source operates at $0.6$\,m depth. Although there is no temporal depth variation, the depth offset among the three entities enhances the observability along the $z$-axis.

\vspace{1mm}
\noindent
\textbf{Hydrophone data collection}. The experiments employ H2dM model hydrophones from Aquarian Hydrophones~\cite{h2dm_hydrophone_misc}. The outputs are terminated with a dual-mono $3.5$\,mm TRS connector, requiring an external plug-in power (PIP) for operation. Therefore, a custom pre-amp circuit is designed to accommodate the input impedance and bias current, as well as to amplify the audio output to a desired voltage range. Both channels of the circuit are identical, ensuring consistent amplification for the two hydrophones. Since the TDOA-FDOA method compares the two waveforms, a higher sampling resolution directly improves the measurement accuracy. Hence, we use a sampling rate of $200$\,kHz, and the two-channel acoustic data are synchronously acquired using a Digilent Analog Discovery (DAD3) board~\cite{dad_board_misc}. 

% \vspace{1mm}
\noindent
\textbf{Acoustic noise filtering}. We identify two dominant noise sources during the experiments: low-frequency water flow noise and high-frequency thruster interference generated by the ASVs. The latter induces an amplitude-modulation effect, resulting in waveform distortion during aggressive maneuvers, while surface bubbling elevates the low-frequency noise floor. To mitigate these effects, each hydrophone signal is processed using a fourth-order zero-phase Butterworth band-pass filter~\cite{butterworth1930theory} centered at the source tone frequency, which is assumed to be close to the data drives' resonant frequency. The zero-phase filtering preserves both temporal alignment and inter-channel delay integrity, ensuring unbiased TDOA estimation. The filtered signals exhibit signal-to-noise ratio (SNR) improvements of up to $40$\,dB, enhancing the subsequent TDOA-FDOA measurements.

%\JI{Needs to be rewritten. First two sentence is basic, the rest reads like a caption to the figures. It is not a results/analysis type of presentation. Same goes for previous few paragraphs as well.}
% Quantitative analysis of the field data highlights the critical influence of filter parameterization and motion modeling on overall localization performance. 
%\Adnan{Revised this para and added details in previous three paras} 

\vspace{1mm}
\noindent
\rebuttal{\textbf{Localization performance}. Real-world acoustic measurements exhibit higher temporal fluctuation than their geometric predictions, as seen from the measured and modeled TDOA-FDOA traces in Fig.~\ref{fig:field_result}\,a. Therefore, we empirically calibrate the process and measurement covariances from $15$ independent sets of field data, and evaluate both the nominal and compensated filtering variants. The nominal filter uses the direct-path measurements, while the compensated filter incorporates the proposed multipath-aware correction and receiver-uncertainty covariance inflation. For the multipath-aware model, the water depth is obtained from a bathymetry chart of the lake site~\cite{lake_wauburg_bathymetry}. The surface and bed reflection coefficients are set to $-0.98$ and $0.2$, respectively, to match first-order approximations~\cite{akal1972relationship,bolghasi2017low}. Fig.~\ref{fig:field_result}\,b shows that both filters track the overall source trajectory, but the compensated estimate remains closer to the true path, especially at the curved segments. The corresponding error plot in Fig.~\ref{fig:field_result}\,c shows a moderate but consistent improvement, reducing the RMSE from $1.89$\,m to $1.37$\,m.
% Moreover, the localizer maintains errors below $2$\,m during smooth trajectory segments, while transient spikes occurring during sharp turns or high angular accelerations are rapidly corrected within a few iterations (see Fig.~\ref{fig:field_result}\,c). 
Overall, the field results confirm that our localization pipeline remains stable in real shallow-water conditions, while multipath and receiver uncertainty compensation further improve tracking accuracy without requiring additional sensing hardware.}

%% file: src/06_Discussion.tex
\section{Discussion: Challenges and Practicalities}\label{sec:discussion}
\rebuttal{The experimental results highlight both the promise and the practical challenges of the proposed framework. In simulation, the LC-MAP initializer improves convergence time and success rate across all source motion models, demonstrating that geometry-aware initialization is critical for joint TDOA-FDOA filtering. In Gazebo-ROS experiments, the pipeline remains stable when integrated with robotic motions. The field trials, however, reveal some gaps between simulation and real-world deployment due to higher-order reflection, limited vertical motion, and acoustic interference from ASV thrusters. This section discusses these practical factors and outlines the scope for future improvements.}

\subsection{TDOA Measurement Challenges}
% \vspace{1mm}
\noindent
\textbf{Standing waves}. While multi-channel synchronization is a well-known challenge in TDOA estimation, we identified an additional and often overlooked issue during our experiments in a $3$\,m$\times$\,$2$\,m$\times$\,$1.5$\,m water tank. Despite varying the baseline distance between two hydrophones, their phase difference remained nearly constant. This phenomenon is attributed to standing wave formation, since the tank's dimension is comparable to signal wavelength ($\approx1$\,m at $1.5$\,kHz). The attack signal swept frequencies between $500$\,Hz and $1.$5\,kHz, which also coincides with the mechanical resonance band of the data center prototype. Within this frequency range, reflections from the tank walls, floor, and surface boundaries create a spatially uniform acoustic phase field and make the phase difference unobservable. We observed small phase variations only when the hydrophones were placed very close to reflective boundaries, such as the tank floor. This occurs since boundary conditions slightly distort the standing pressure field, producing small, local phase gradients. The observations provide two key lessons for acoustic experiment design: \textbf{(i)} even mild wall reflections can alter phase/delay relationships across channels, so apparent clean signals may not be physically meaningful for delay-critical tests; \textbf{(ii)} more suitable environments for such tests are large open-water areas or tanks equipped with acoustic absorbers to suppress reflections.

\vspace{1mm}
\noindent
\rebuttal{\textbf{Multipath interference}. Multipath propagation is known to perturb phase-difference and TDOA measurements in reverberant underwater environments~\cite{liu2023efficient}. We incorporate a lightweight multipath-aware correction model by considering two dominant reflected paths: the water-surface reflection and the bed reflection. As shown in Sec.~\ref{subsec:multipath_ablation}, this correction helps the filter to converge and reduces the average RMSE from $25.71$\,m to $0.98$\,m in simulation tests. Despite this improvement, the model does not account for higher-order reflections, scattering from the data center pod and from the mobile robot's body, or other time-varying environmental effects. Moreover, the LC-MAP cannot benefit from multipath correction since it requires an estimate of the source position. During initialization, no prior knowledge of the source is known. Hence, LC-MAP estimates a physically plausible initial state using only observed TDOA-FDOA. Subsequently, the UKF refines the estimate using the multipath-aware measurement model. Incorporating multipath information directly into the initialization stage is a promising extension, which requires either a precomputed acoustic field model or prior knowledge of the source location.
}

\vspace{2mm}
\noindent
\rebuttal{\textbf{GCC-PHAT limitation under narrowband phase ambiguity}. Generalized cross correlation with phase-transform (GCC-PHAT)~\cite{knapp1976generalized} is a widely used tool for TDOA estimation in broadband signals. However, its effectiveness is limited for narrowband or tonal emissions~\cite{salvati2021time,yang2025neural}. For such signals, the correlation response becomes periodic, producing multiple plausible delay candidates separated by the signal period. Thus, GCC-PHAT does not resolve the fundamental phase-wrapping ambiguity inherent to tonal signals. Hence, we use short-window cross-correlation followed by temporal unwrapping to estimate temporally consistent TDOAs in dynamic transmitter-receiver settings.}

% Our prior acoustic adversarial analyses were conducted in a tank testbed~\cite{blow2025detection,sheldon2024aquasonic}, which we found unsuitable for this study. Since the data center prototype's structural resonance frequency is approximately $1.5$\,kHz, any signal near this frequency band creates standing pressure waves, resulting in a nearly uniform acoustic phase field. As a result, the phase differences between hydrophone signals -- essential for accurate TDOA estimation -- become unobservable except very close to the reflective surfaces (\eg, tank floor). Essentially, the tank testbed prevents valid TDOA/FDOA analyses, and all localization experiments in this work are conducted in open-water conditions instead.

\subsection{Challenges in Real-world Tests}
\rebuttal{The shallow-water field trials revealed several practical factors that explain the performance gap between controlled simulations and real-world deployment.} First, the mobile hydrophone was mounted beneath a BlueBoat~\texttrademark{} ASV, restricting its motion to a near-planar trajectory at a fixed depth. Consequently, spatial diversity along the vertical ($z$) axis was limited and occasionally led to poor geometric conditioning in the UKF's state estimates~\cite{reis2018source}. To mitigate this issue, the vertical position uncertainty was tightly bounded, and the depth was initialized close to its true value. Second, the ASV's onboard GPS drifted up to $2$\,m, introducing bias in both the ground-truth trajectory and the geometric modeling of TDOA/FDOA. We partially compensated for this effect by inflating the effective measurement covariance based on mobile receiver state uncertainty. We are integrating more accurate navigation aids, such as RTK-GPS, for centimeter-level positioning accuracy. Additionally, for GPS-denied underwater deployment, the surveillance robot will rely on visual-inertial odometry and SLAM to maintain its pose with respect to the data center pod.

Third, the hydrophone signals were affected by acoustic interference from surface waves, air bubbles, and thruster noise, which raised the background spectral floor and occasionally overlapped with the attack signal band. The interference was suppressed using a narrow bandpass filter centered on the source tone; however, the potential spectral overlap remains a limitation in low-SNR conditions. \rebuttal{Finally, our shallow-water multipath compensation relied on approximate environmental parameters. In particular, we inferred water depth from a bathymetry chart~\cite{lake_wauburg_bathymetry} and used nominal values for the reflection coefficients rather than in-situ calibration. Even after this first-order modeling, residual errors remain due to unmodeled pod reflections, higher-order multipath, surface roughness, and imperfect reflection coefficients.}

\rebuttal{Overall, these trials show that the real-world performance of TDOA/FDOA-based localization is bounded by four critical factors: \textbf{(i)} geometric diversity of receiver motion, \textbf{(ii)} positional accuracy of the mobile receiver, \textbf{(iii)} acoustic signal quality under environmental interference, and \textbf{(iv)} the accuracy of the multipath model. Addressing these factors, particularly via precise and diverse receiver motion, will further enhance localization and tracking accuracy.}

\subsection{Improvements and Future Works}

\vspace{1mm}
\noindent
\textbf{Deep water deployment with submerged UDC pod}. We are developing a new platform to evaluate adversarial acoustic localization in fully submerged environments. A custom-built data center pod will be deployed at $20$-$30$\,m deep open water lake and grotto systems to mimic realistic subsea operating conditions. Consequently, two custom AUVs, CavePI~\cite{gupta2025demonstrating} and NemeSys~\cite{abdullah2025nemesys}, will replace the BlueBoat~\texttrademark{} ASVs previously used as the adversarial and surveillance agents. In the absence of GPS, the AUVs will use fiducial markers (mounted on the pod's outer surface) and multi-sensor SLAM~\cite{rahman2022svin2} for self-localization and navigation. The pod will house multiple server units and remain tether-connected to an on-land operations hub for remote command. During a controlled attack scenario, this testbed will support an end-to-end demonstration from attack detection and localization to coordinated mitigation actions such as traffic isolation, data replication, threat neutralization, etc.

% This setup will facilitate simultaneous self-pose estimation and adversarial source localization.

\vspace{1mm}
\noindent
\textbf{Observability-driven active surveillance path planning}. Utilizing two hydrophones for 3D source localization is inherently an under-constrained problem. We overcome this limitation by mobilizing one hydrophone and collecting diverse spatial observations over time. The surveillance robot's trajectory determines this observability and thus affects the filter's convergence. In our experiments, purely linear motions yield poor geometry and near-singular Jacobians, which occasionally lead to estimator divergence. %\Adnan{revised from here:} 
LC-MAP mitigates this at initialization by selecting well-conditioned priors. Future work will extend these principles to an adaptive path-planning framework. For instance, an \textit{observability-aware} path planner would favor paths that maximize measurement diversity and minimize estimation uncertainty. Such adaptive guidance is expected to further improve localization accuracy under weak-geometry conditions.
%\textcolor{red}{Sara: one consideration of this, if you say it like this it seem that we know there are better methods (more complete long-period monitoring) but we did not implement those (which sounds like the paper is not complete). You might want to rephrase that to describe the advantages of the current technique and that for future work might take in consideration to continuos path}

\vspace{1mm}
\noindent
\rebuttal{\textbf{Matched-field processing for multipath effect mitigation}. While the current UKF model compensates for first-order surface and bed reflections, future work will extend toward matched-field processing for improved multipath modeling~\cite{chen2012source}. Since the data center facility is static and the surrounding environment can be mapped once during deployment, an acoustic replica field can be constructed for the surveillance region. During an acoustic attack, matched-field methods can compare the received acoustic field against a precomputed replica and provide additional spatial constraints for localization. This is particularly suitable for UDC surveillance because the infrastructure geometry is fixed and can be calibrated over time, while the mobile receiver can actively sample informative viewpoints to improve the acoustic field model.
}

% \vspace{1mm}
% \noindent
% \rebuttal{\textbf{Hybrid sensing and learning-based extensions}. The proposed acoustic framework can be integrated with complementary sensing modalities such as imaging sonar, optical cameras, and underwater LiDAR when environmental conditions permit. In addition, learning-based acoustic localization may become feasible once sufficiently large labeled datasets are available across source positions, receiver trajectories, multipath conditions, and attack frequencies. A promising direction is hybrid model-learning, where TDOA-FDOA acoustic measurements provide geometric constraints while learned models compensate for environment-specific acoustic distortions.}

%% file: src/07_Conc.tex
\section{Concluding Remarks}\label{sec:conclusion}
This work focuses on localizing adversarial acoustic agents that threaten offshore data centers, where direct human supervision is not feasible. 
% The proposed system focuses on HDD-targeted acoustic injection attacks, where sound can mechanically perturb storage devices and degrade I/O availability.
% The proposed framework provides a real-time solution for detecting and localizing acoustic threats in UDCs through continuous monitoring and adaptive signal analysis. 
By estimating the adversarial source position and motion, this work enables rapid defensive intervention, effectively mitigating the risk of data corruption and hardware damage from targeted acoustic injections.
The proposed system uses a minimal yet scalable heterogeneous receiver configuration, and thus eliminates the logistical overhead of dense, synchronized arrays while retaining full 3D localization capability. Unlike conventional TDOA-FDOA localizers that rely on ad-hoc initialization and fail under uninformed, stealthy target conditions, our proposed LC-MAP scheme leverages acoustic geometry to maximize observability during the early measurement stage. 
\rebuttal{We further incorporate a multipath-aware TDOA model to account for first-order surface and bed reflections, as well as an effective measurement covariance that compensates mobile receiver state uncertainty.} Integrated into a UKF pipeline, the framework enables fast and robust 3D localization of dynamic adversarial sources under noisy measurement conditions. \rebuttal{Across Monte Carlo simulations with four source motion models, the proposed method achieves over $90\%$ success rate in most scenarios and reduces average convergence time by nearly $50\%$ compared to the naive baseline. In Gazebo-ROS physics simulations, the pipeline achieves sub-meter position accuracy, while open-water field trials demonstrate practical feasibility under real-world noisy measurements.}
% Extensive numerical analyses, Gazebo-ROS physics simulations, and open-water field trials confirm that the proposed LC-MAP-integrated pipeline achieves sub-meter localization accuracy and reduces convergence time by nearly half compared to off-the-shelf state estimation filters. 
Beyond data center security, the proposed framework establishes a scalable, generalizable foundation for acoustic surveillance of offshore assets, such as underwater sensor networks, communication relays, and distributed marine infrastructure.

\section*{Acknowledgments}
\vspace{-2mm}
This research is supported by the Office of Naval Research (ONR) grant N000142412596. 

% \section*{Supplemental Materials}
% \rebuttal{A video demonstration of the system is available here: \url{https://youtu.be/6QOY7q3n34M}.}